\documentclass[11pt]{article}
\usepackage{graphicx,floatflt,amssymb} 
\usepackage{epsfig} 
\usepackage{graphics} 
\usepackage{psfrag}
\input epsf
\def\beq{\begin{equation}}
\def\eeq{\end{equation}}
\def\bea{\begin{eqnarray}}
\def\eea{\end{eqnarray}}
\def\bq{\begin{quote}}
\def\eq{\end{quote}}
\def \lsim{\mathrel{\vcenter
     {\hbox{$<$}\nointerlineskip\hbox{$\sim$}}}}
\def \gsim{\mathrel{\vcenter
     {\hbox{$>$}\nointerlineskip\hbox{$\sim$}}}}
\def\gappeq{\mathrel{\rlap {\raise.5ex\hbox{$>$}}
{\lower.5ex\hbox{$\sim$}}}}
\def\lappeq{\mathrel{\rlap{\raise.5ex\hbox{$<$}}
{\lower.5ex\hbox{$\sim$}}}}

\def\bea{\begin{eqnarray}}   
\def\eea{\end{eqnarray}}
\def\la{\mathrel{\mathchoice {\vcenter{\offinterlineskip\halign{\hfil
$\displaystyle##$\hfil\cr<\cr\sim\cr}}}
{\vcenter{\offinterlineskip\halign{\hfil$\textstyle##
$\hfil\cr<\cr\sim\cr}}}
{\vcenter{\offinterlineskip\halign{\hfil$\scriptstyle##
$\hfil\cr<\cr\sim\cr}}}
{\vcenter{\offinterlineskip\halign{\hfil$\scriptscriptstyle##$
\hfil\cr<\cr\sim\cr}}}}}
\def\ga{\mathrel{\mathchoice {\vcenter{\offinterlineskip\halign{\hfil
$\displaystyle##$\hfil\cr>\cr\sim\cr}}}
{\vcenter{\offinterlineskip\halign{\hfil$\textstyle##
$\hfil\cr>\cr\sim\cr}}}
{\vcenter{\offinterlineskip\halign{\hfil$\scriptstyle##
$\hfil\cr>\cr\sim\cr}}}
{\vcenter{\offinterlineskip\halign{\hfil$\scriptscriptstyle##$
\hfil\cr>\cr\sim\cr}}}}}

\begin{document}
\vspace*{-1in}
\renewcommand{\thefootnote}{\fnsymbol{footnote}}
\begin{flushright}
\texttt{CERN-PH-TH/2006-001}\\
\texttt{DNI-UAN/06- 01FT}\\
\texttt{LPT-Orsay/06-03}\\
\texttt{LYCEN 2006-03}\\
%\texttt{hep-ph/06????} 
\end{flushright}
\vskip 5pt
\begin{center}
{\Large {\bf Flavour Issues in Leptogenesis
}}
\vskip 25pt
{\bf Asmaa Abada $^{1,}$\footnote{E-mail address:
abada@th.u-psud.fr}, Sacha Davidson $^{2,}$\footnote{E-mail address:s.davidson@ipnl.in2p3.fr},
Francois-Xavier Josse-Michaux$^{1,}$\footnote{E-mail address:
josse@th.u-psud.fr},\\
 Marta Losada $^{3,}$\footnote{E-mail address:
malosada@uan.edu.co}   and Antonio Riotto$^{4,}$\footnote{E-mail address: 
antonio.riotto@pd.infn.it}}
 
\vskip 10pt  
$^1${\it Laboratoire de Physique Th\'eorique, 
Universit\'e de Paris-Sud 11, B\^atiment 210,\\ 91405 Orsay Cedex,
France} \\
$^2${\it Institut de Physique Nucl\'eaire de Lyon, Universit\'e C-B Lyon-1,
69622 Villeurbanne, cedex,  France }\\
$^3${\it Centro de Investigaciones, 
Universidad Antonio Nari\~{n}o, \\ Cll. 58A No. 37-94, Santa Fe de Bogot\'{a},
Colombia}\\
$^4${\it CERN Theory Division, Geneve 23, CH-1211, Switzerland}

\vskip 20pt
\vskip 20pt {\bf Abstract}
\end{center} 
 
\begin{quotation} {\noindent
We study the impact of  flavour in thermal leptogenesis, including the
quantum oscillations of the asymmetries in lepton flavour space.
In the Boltzmann equations we 
find different numerical factors and
additional terms which can affect the results significantly.
The upper bound on the CP asymmetry in a specific flavour
is weaker than the bound on the sum. 
This suggests that
-- when flavour dynamics is included --
there is no model-independent limit on the light neutrino 
mass scale,
and that 
the lower 
bound on the reheat temperature is relaxed by 
 a factor $\sim (3 - 10)$.
 
\vskip 10pt 
\noindent PACS number(s):~12.60.Jv, 14.60.Pq, 11.30.Fs\\
}

\end{quotation} \noindent{Leptogenesis, Neutrino Physics, Baryon Asymmetry of
the Universe}

\vskip 20pt  

\setcounter{footnote}{0} \renewcommand{\thefootnote}{\arabic{footnote}}

%%%%%%%%%%%%%%%%%%%%%%%%%%%%%%%%%%%%%%%%%%%%%%%%%%%%%%%%%%%%%%%%%%%%%% 
%%Introduction %%%%%%%%%%%%%%%%%%%%%%%%%%%%%%%%%%%%%%%%%%%%%%%%%%%%%%
%%%%%%%%%%%%%%%%%%%%%%%%%%%%%%%%%%%%%%%%%%%%%%%%%%%%%%%%%%%%%%%%%%%%%%

% \newpage

{\bf Note added:} The  equations (\ref{osc}) and (\ref{compact}) of
the published paper, 
which include the decaying oscillations in flavour space, do not
 reflect the flavour structure derived in the Appendix,
or reproduce the behaviour discussed in the text.  
  In  this version, we modify  equations
 (\ref{osc}) and (\ref{compact}), and  add a discussion, in
Appendix A, on how to obtain these improved equations. 
We thank Georg Raffelt for drawing our attention to
this, (and also to the paper by N.F.Bell {\it et al,
Phys. Lett.} {\bf B500} (2001) 16, which indicates that
fast gauge interactions should not affect the coherence of flavour
oscillations). We also remove the errant factor of $z$
multiplying the  charged lepton Yukawa terms of  
(\ref{osc}) and (\ref{compact}).  
Equations (\ref{l},\ref{asymf}) are solutions to the flavour
structure of the Appendix, so  should be 
 unaffected by the change of form of
equations (\ref{osc}) and (\ref{compact}).
However, (\ref{l},\ref{asymf}) are   
 are wrong by the factor $z$.

\section{Introduction}

\noindent
Leptogenesis \cite{FY} is 
an attractive mechanism to 
explain the baryon asymmetry
of the Universe (BAU) \cite{sakharov}. It consists of the dynamical production of a lepton asymmetry in
the context of a given model, which then can be converted to a baryon asymmetry
due to $(B+L)$-violating sphaleron interactions \cite{kuzmin}
which exist in the Standard Model (SM).

A simple model in which this mechanism can be implemented is ``Seesaw''(type I)
\cite{seesaw}, consisting
of the Standard Model (SM)
plus 3 right-handed (RH) Majorana neutrinos.
In this simple extension of the SM, the usual scenario that is explored consists of
a hierarchical spectrum for the RH neutrinos, such that the 
lightest of the RH neutrinos
is produced by thermal scattering after inflation,  and
decays out-of-equilibrium in a lepton number and CP-violating way, thus satisfying Sakharov's constraints.  In recent years, a lot of 
work \cite{leptogen,work}, 
has been devoted to a thorough analysis of this model, the constraints that can be imposed,
the relationships that arise with low energy neutrino physics, etc.
For instance,  a lower bound on  the reheating temperature 
$T_{\rm RH}$ and an upper bound on
the scale of light neutrino masses can be obtained \cite{di,bound,bdp2,HLNPS}.

However, in order to obtain an accurate value of the produced lepton asymmetry it is necessary to
solve (in a first approach at least) 
the system of coupled Boltzmann equations (BE) which
relate the abundance of the lightest RH neutrino with the  asymmetry produced 
in lepton number. The effects of  the charged lepton Yukawa coupling matrix $[h]$
are usually not considered. In this study, we take into account these effects 
 in deriving  the evolution equations  for the asymmetries in each flavour,
and study the implications for leptogenesis.
 Flavour effects  have been also discussed in \cite{barbieri}. More 
recently, the generation of the family lepton 
asymmetries has been discussed   in \cite{kor1} and \cite{kor2}, where
the case of leptogenesis in the minimal seesaw
model with two heavy Majorana neutrinos was considered; the role of
flavours in the leptogenesis scenario   
from $N_2$ decay in discussed in \cite{oscar}
and for ``resonant leptogenesis'' (degenerate
$N_J$)  in \cite{PU}.
  In this work, we find three effects: a qualitative difference
in the equations, which allows quantum oscillations of
the asymmetry in flavour space. We  naively 
anticipate this could change the final asymmetry by a factor $\sim 2$
\cite{todo}. Secondly, we find different numerical factors
compared to the case when no flavour issues are included 
in the Boltzmann equations, and additional terms 
when the flavour-dependent decay rates and
asymmetries are unequal. 
Finally,
  we find  that the CP asymmetries
in individual flavours can be larger than the sum, 
which affects  the leptogenesis bounds on
$T_{\rm RH}$ and makes the bound on the light neutrino mass scale
disappear.

In section \ref{notn} we review our notation and 
summarise the usual way the calculation is performed. 
In section \ref{rho}, we  present
equations  for 
 the lepton asymmetry,  represented as a matrix in flavour space:
the diagonal elements are the  flavour asymmetries, and the off-diagonals
encode quantum correlations. 
These could be relevant, because, as will be explained, the muon Yukawa coupling comes into thermal equilibrium around $T \sim 10^9$ GeV, 
allowing the possibility that the asymmetry  ``oscillates'' in flavour space. 
To motivate the equations 
used in section \ref{rho} we study a quantum mechanical toy model of simple
harmonic oscillators. This study is explained  in appendix A of  this note.
We concentrate on the BE for
the three flavour asymmetries in section \ref{flavour},
 and compare the resulting
equations for the lepton asymmetry with the usual approximation.
We present results in section \ref{applns},
which aims to be comparatively self-contained. Appendix B contains a brief discussion on the $\Delta L =2$ terms, and Appendix C relates
the asymmetries in $B/3 - L_i$ to those in the lepton doublets.

\section{Notation, Review  and the ``Standard'' Calculation}
\label{notn}

We  consider the Lagrangian for the lepton sector 
\beq
{\cal L}= h^{ ik } H \bar{e}_{R i} \ell_k + 
\lambda^{Jj} H^c \bar{N}_J \ell_j + 
\frac{M_{K}}{2} N_{K} N_{K} + {\rm h.c.}
\label{L}
\eeq
in the  mass eigenstate bases  of  the singlet (``right-handed'')
 particles $N$ and  $e_R$. The  mass eigenstate basis for
the doublet leptons will be temperature dependent, so we do
not fix it here. 
We assume from now on a hierarchical spectrum for the RH neutrinos, $M_{1} \ll M_{2}< M_{3}$.

The light neutrino mass matrix, in the charged lepton mass
eigenstate basis, is
\beq
m_\nu   = U^* D_{m_\nu} U^\dagger  =  \lambda^T M^{-1} \lambda v^2 \ ,
\eeq
where $U$ is the PMNS matrix, $D_A$ is a diagonal matrix 
of real eigenvalues of $A$ and $v$ the Higgs vacuum expectation value.

The physical process of interest is the production of a lepton asymmetry in
the decay of  the lightest  singlet neutrino $N_1$, in the thermalised early
Universe.  
  We make the usual assumption, that
interactions  that are much faster
than  the expansion rate $H$ --- such as gauge interactions,
$(B+L)$-violation,  and (possibly) the
$\tau$-Yukawa coupling ---  are  in  equilibrium, which imposes
certain conditions  \cite{HT} on the  distribution functions of particles.
 For instance \cite{CDEO,CKO}, when  
\beq
\Gamma_{\tau} \sim
5 \times 10^{-3} h_\tau^2 T\gg H
\label{yuk}
\eeq
the chemical potentials of the singlet
and doublet $\tau$s satisfy the relation 
%for the corresponding  chemical potentials
 $\mu_H + \mu_{\tau_L} = \mu_{\tau_R}$. 
We neglect interactions whose rates are
much smaller than $H$, {\it e.g.} those induced by the electron Yukawa
coupling.
Interactions that are of order the
expansion rate, and those that
are lepton number violating, are included
in the evolution equations for the number density (eventually,
Boltzmann equations).    These  are
 the $N_1$-interactions,
the $\Delta L=1 $ and $\Delta L = 2$ scatterings, 
and possibly the interactions with strength 
set by the $\mu$-Yukawa coupling, $h_{\mu}$.

It is  straightforward 
to write down Boltzmann equations  for the conserved
asymmetries of  the Standard Model (SM). The SM interactions
which are in equilibrium define the conserved quantum numbers.
For instance, below $T \sim 10^5$ GeV, when all interactions
are in equilibrium,  the conserved quantum numbers are
the   $\{ B/3 - L_i \}$ for $i=e,\mu,\tau$.
The Boltzmann equations are obtained by
identifying processes that change 
the asymmetries, and  summing their rates. The only
delicate point arises
in the temperature range where the interactions induced by 
$h_\mu$ come into equilibrium, 
because this changes the relation between $Y_{L_\mu}$ and
 $Y_{B/3 - L_\mu}$.

The Boltzmann equations for number densities  
neglect the possibility of quantum effects,
such as oscillations, due to  interference
among different processes.  Such effects could appear in
 the equations of motion
for  the (flavour-dependent)
 number operator \bea\label{fdelta}
\hat{f}_\Delta^{i j} (\vec{p}) 
=   a_+^{i \dagger}(\vec{p})  a^j_+ (\vec{p})  -
 a_-^{j \dagger}(\vec{p})  a^i_-(\vec{p}),\eea
which counts  the asymmetry in lepton doublets\footnote{
$\hat{f}$ wears a hat as an operator, to distinguish
it from its expectation value which we will usually
discuss in the Appendix.
The operator $ a_+^{i \dagger}$
$( a_-^{i \dagger})$ creates 
particles (anti-particles)
of  flavour $i$.}. Notice the inverted flavour order
between the particle and antiparticle number operators
\cite{RS}. 
The operator is a matrix
in flavour space, analogous to
the density matrix of quantum mechanics,
and is sometimes called a density matrix.
 When the flavour indices
$ij$ are the charged lepton mass eigenstates, 
then the diagonal elements  are the  flavour asymmetries
stored in the lepton doublets, and should satisfy Boltzmann(-like) Equations.
The trace, which is flavour-basis-independent, is the total 
lepton asymmetry. The off-diagonals encode
(quantum) correlations between the different flavour asymmetries,
and should decay away when the charged Yukawa couplings are in equilibrium
\cite{barbieri}. 
Variants on such an operator have been
studied in  the context of
neutrino oscillations in the early
Universe \cite{RS,leo,McKellar,Raffelt:1992uj},
and in particular the generation of a lepton asymmetry
by active sterile oscillations
\cite{Enqvist:1990ad,Foot:1996qc}.
This operator makes  brief appearances in \cite{barbieri}
 where the issue of flavour in 
the context of thermal leptogenesis
was  first discussed.

In appendix \ref{apenrho}, we will  motivate  equations of motion 
for
these flavour-dependent  lepton number operators 
using  quantum mechanical oscillators.  For simplicity,
we separate  the  (possibly quantum) flavour effects from
the (assumed classical) particle dynamics.  That is,  we extract 
the flavour structure from the toy model, and input the particle dynamics
and the Universe expansion,  by
analogy with the Boltzmann equations. This is an
elegant way to obtain flavour-dependent
Boltzmann equations, because the tensor
structure of the operators in flavour space is clear. In a
later publication, we will derive
equations of motion for  $f_\Delta^{ij}$ in
field theory. Indeed, 
its dynamics 
could  be obtained from the transport equations deduced from the 
finite temperature  Schwinger-Dyson equations
involving flavoured fermionic propagators. In this set-up \cite{ar}, 
one makes use of the 
 closed time-path formalism to describe nonequilibrium phenomena in 
field theory leading to a complete nonequilibrium quantum kinetic theory. 
The operators $f_\Delta^{ij}$ are  associated 
to propagators which may mix the flavour and which are ``sourced'' in the
Schwinger-Dyson equations by the 
flavour-dependent CP asymmetry  $\epsilon_{ij}$ from
the decay of the right-handed neutrino.

The usual starting point for leptogenesis is a system of
coupled equations.  If the asymmetry is due
   to the decay of the lightest
    RH neutrino $N_1$,
 and the processes included  are decays, inverse decays  
and  lepton number violating scattering ($\Delta L=1$
from Higgs exchange, and  $\Delta L=2$ due
to  RH neutrinos), then these equations are:

\beq
\frac{dY_{N_{1}}}{dz} = - \frac{z}{s H(M_1)} \left( \gamma_D + \gamma_{\Delta L =1}\right)
 \left(\frac{Y_{N_{1}}}{Y_{N_{1}^{eq}}} -1\right),
\label{YN1s}
\eeq

\beq
\frac{dY_L}{dz}  =    \frac{z}{s H(M_1)}\left[\left(
\frac{Y_{N_{1}}}{Y_{N_{1}}^{eq}} - 1\right)\epsilon \gamma_D -
\frac{Y_L}{Y_L^{eq}} \left(\gamma_D
+\gamma_{\Delta L =1} + \gamma_{\Delta L=2}\right)\right].
\label{YLs}
\eeq
 $Y_{N_{1}}$ is the abundance of the  RH neutrino $N_{1}$ 
normalized to the entropy density $s$, and 
$Y_L$ is the sum over flavour of the difference between normalized 
abundances of leptons and antileptons.
In eqs. (\ref{YN1s}) and (\ref{YLs}),  $Y^{eq}_{i}$ is the equilibrium number 
  density of
  a particle $i$, $z=\frac{M_1}{T}$ and the quantities  $ \gamma_D,   \gamma_{\Delta L=1}$ and 
$\gamma_{\Delta L =2}$ 
are thermally averaged rates
of decays, inverse decays, and scattering processes 
with $\Delta L= 1,2$ 
and include all contributions
summed over flavour ($s$, $t$ channel
interference etc). The explicit expressions are
given in the literature (see for example   \cite{leptogen}) and we will discuss them when needed further on. 
 The total
CP-asymmetry  $\epsilon$ produced in
the decay of the lightest RH neutrino is given by
\cite{roulet}
\footnote{We exploit this expression in the next section.}\bea 
\epsilon = \frac{1}{8\pi [\lambda \lambda^\dag]_{11}}
 \sum_{j\neq 1} {\Im}{ [\lambda \lambda^\dag]_{1J}^2}
 f(M_{{J}}^2/M_{{1}}^{2}),
\label{eps1}
\eea
 where $f$ is the loop factor \cite{roulet}, given by  
\bea f(x) &=& \sqrt x \left[1 - (1+x)\ln\frac{1+x}{x} + \frac{1}{1-x}
\right]. 
\label{floop}
\eea  
We recall that the rates $\gamma_{D}$ \footnote{For a later use, we  give the expression of the reaction density $\gamma_{D}=n^{eq}_{N_1}{\frac{K_1(z)}{K_2(z)}}\Gamma,$
where  $K_n(z)$ are the $K$ type Bessel functions and $n_{N_1}^{eq}$ is the equilibrium density.}  and $\gamma_{\Delta L= 1}$  are both proportional to the tree-level decay rate   given by
\bea  \Gamma=\frac{\left(\lambda\lambda^\dagger\right)_{11}}{8\pi}{M_1}=
{\tilde {m}_{1}M_1^2\over 8\pi v^2},\label{gammaN} 
\eea
and the out of equilibrium condition is satisfied  when: \beq\label{outofeq}
K\equiv\left. \frac{\Gamma}{H}\right|_{T=M_1}=\left(\frac{\tilde{m}_1}{
\tilde{m}_*}\right)< 1.\eeq
In other words, comparing $\tilde{m}_1$ to 
$\tilde{m}_{*}\sim 2.3 \times 10^{-3}$ eV 
gives a measure of whether the decay is 
occurring out of equilibrium or not.
 The rescaled decay rate  $\tilde{m}_1$
is bounded below: $m_1\leq \tilde{m}_1 $,
and naturally of order $\sqrt{\Delta m^2_{atm}}$
when the light neutrino spectrum is hierarchical
\cite{Buchmuller:2003gz}.

The term $\gamma_{\Delta L=2}$\footnote{We emphasize 
that  $-{1\over 8} \gamma_{_{D_j}}$ corresponding to RIS (real
intermediate states) in the $\Delta L=2$ interactions have 
to be carefully subtracted, to avoid
 double counting in the Boltzmann equations \cite{leptogen}.} 
has a more complicated dependence on the 
masses of the neutrinos, and can be treated in two separate regimes. 
The contribution in the temperature range from $z=1$  to $z\simeq 10$
is proportional to $\tilde{m}_1$ while for the range of
temperature $(z\gg 1)$, the dominant contribution at leading order 
is proportional to 
 $\bar{m}^2 \equiv {\rm Tr}(m_{\nu}^{\dagger}m_{\nu})=\sum_i m_i^2$. 
Here $m_i$ denotes the mass eigenvalues of the left-handed neutrinos.

\section{Flavour Oscillations}
\label{rho}

In this section we  consider thermal 
leptogenesis for  two flavours,
which to be explicit, we label to be  the $e$- and the 
$\mu$ leptons. This is  interesting  in two cases.
First, if leptogenesis takes place  when $\Gamma_\mu \ll H$
(see eq. (\ref{yuk})),
%well before $h_\mu$ comes
%into equilibrium, 
it will be described by Boltzmann
equations for the $\tau$ and $\ell_1'$
number densities, where $\vec{\ell}_1' = \vec{\ell}_1 - (\vec{\ell}_1 \cdot 
\hat{\tau}) \hat{\tau} $ and $({\ell}_1)_i   =
\lambda_{1 i}$  is the decay direction,
in lepton flavour space,  of $N_1$. 
So the equations of this section could
be applied in this high temperature case, 
with the relabelling $\mu \rightarrow \tau$,
$e \rightarrow \ell_1'$.
Secondly, if the muon Yukawa coupling comes into equilibrium as
the asymmetry is being created,
there  could be   oscillations of the asymmetry, parametrised in
off-diagonal elements of the asymmetry density.
Focusing on this quantum effect,  we neglect
the third lepton generation, because  we suppose
we are in the
interesting range of temperatures  $ \lsim 10^{12}$ GeV
where
 the interactions  induced
by the $\tau$-Yukawa coupling are in chemical equilibrium. 
 Therefore  no quantum correlations
between the lepton asymmetries involving the third family are expected to 
survive.
Adding the third family will be discussed in the next section.

The equations describing the two-flavour system have been obtained
in appendix \ref{apenrho}, to which  we refer 
the reader for more details. To
clearly see the effect on flavour oscillations due
to the muon Yukawa coupling coming into equilibrium, we map 
eq.  (\ref{damp})  onto  two-flavour equations
for the asymmetry in  the early Universe, 
  neglecting
the transformation to $(B/3 - L_i)$.
 As discussed after eq. (\ref{damp}), these
equations describe decays and inverse
decays of $N_1$, and the resonant part
of $\Delta L = 2$ scattering is also included.
 We work in the charged lepton mass
eigenstate basis, and distribute  indices by
analogy with the model of the appendix 
($f_{\Delta \ell}^{ij} \leftrightarrow 
Y_L^{ij}$,  and $\lambda^{i}\lambda^{* j} 
\leftrightarrow  \gamma^{ij}$). 

We assume, to obtain these equations, that the flavour-blind
gauge interactions  can be neglected. However,
the oscillation frequency  in flavour space 
depends on the energy of the lepton, and
 within the oscillation timescale, a lepton will participate in many
energy-changing gauge interactions\footnote{We thank
A. Strumia for discussions about this point.}.
In this note,  we  use the thermally averaged energy $\langle
E \rangle$  to
estimate the oscillation frequency \cite{Enqvist:1990ad}. That is, we
approximate the integral $ (i \int E d \tau)$ 
along the path from one  lepton-number violating
interaction to the next, to be 
$i \langle
E \rangle \int  d \tau$.  
 We will
include the gauge interactions more correctly in a
later analysis \cite{todo}.

The system describing the two flavours is given by 

%{\footnotesize
\bea
\frac{d}{dz} 
\left[
\begin{array}{cc}
Y_{L}^{ee} &Y_{L}^{e \mu}  \\
Y_{L}^{ \mu e}&Y_{L}^{ \mu \mu}
\end{array}
\right]
& = & \frac{z}{s H(M_1)}\left(\gamma_D 
\left( \frac{Y_{N_{1}}}{Y_{N_{1}}^{eq}} -1 \right)
\left[
\begin{array}{cc}
\epsilon^{ee} &\epsilon^{e\mu}  \\
\epsilon^{ \mu e}&\epsilon^{\mu \mu}
\end{array}
\right] 
\right.
% \nonumber \\
%&-& 
- \left. 
 \frac{1}{2 Y_{L}^{eq}}\left\{ 
\left[
\begin{array}{cc}
\gamma ^{ee}_D &\gamma ^{e\mu}_D  \\
\gamma ^{\mu e}_D&\gamma ^{\mu \mu }_D
\end{array}
\right],
\left[
\begin{array}{cc}
Y_{L}^{ee} &Y_{L}^{e \mu}  \\
Y_{L}^{ \mu e}&Y_{L}^{ \mu \mu}
\end{array}
\right]
 \right\}\right)
\nonumber \\
&&
 -i \left[ 
[\Lambda_\omega],
\left[
\begin{array}{cc}
Y_{L}^{ee} &Y_{L}^{e \mu} \\
Y_{L}^{ \mu e}&Y_{L}^{ \mu \mu}
\end{array}
\right]
 \right]
 - \,\frac{1}{2}\left\{ 
[\Lambda_d],
\left[
\begin{array}{cc}
Y_{L}^{ee} &Y_{L}^{e \mu} \\
Y_{L}^{ \mu e}&Y_{L}^{ \mu \mu}
\end{array}
\right]
 \right\}+  [\Lambda_d] Y_{L}^{ \mu \mu}
\label{osc}
\eea
%}
where 
 the parenthesis $\left[\cdot,\cdot\right]$ and $\left\{\cdot,\cdot
\right\}$ stand for commutators and anti-commutators, respectively.
The matrix $[\gamma_D]$ is defined as
\beq
\gamma^{ij}_D = \gamma_D \frac{\lambda_{1i} 
\lambda_{1j}^*}{\sum_k|\lambda_{1k}|^2} , ~~~
\label{gij}
\eeq
where  $ \gamma_D  = \sum_i \gamma^{ii}_D$ is the total 
(thermally averaged) decay rate.
For $i = j$, 
$ \gamma^{ii}_D$ is the  decay rate of $N_1$ to the $i$-th flavour.
The CP asymmetry in the $i$-th flavour  is  $\epsilon^{ii}$,
$\epsilon = \sum_i  \epsilon^{ii}$, and  
notice that $ \epsilon^{ij}$  is  normalised
by the total decay rate  (so that $[\epsilon]$
transforms as a tensor under $\ell$ basis
rotations). The matrix $[\epsilon]$  is now defined as 
\bea
\epsilon_{ ij} &  = &
\frac{1}{(16\pi)}\frac{1}{  [\lambda \lambda^{\dagger}]_{11}}
\sum_J \Im \left\{ ( \lambda_{1 i} ) [ \lambda
\lambda^{\dagger}]_{1J}
 \lambda^*_{J j} - 
 ( \lambda^*_{1 j} ) [ \lambda^*
\lambda^{T}]_{1J}
 \lambda_{J i} \right\}
\left[f\left(\frac{M_{J}^2}{M_{I}^{2}}\right)\right]
\nonumber \\
&\simeq&  \frac{3}{(16\pi v^2)}
\frac{M_1}{  [\lambda  \lambda^{\dagger}]_{11}}
 {\rm Im}  \{\lambda_{i} \lambda_\sigma   [m_\nu^*]_{ \sigma j} 
-  \lambda^{*}_{ \sigma} [m_\nu]_{\sigma i }  \lambda^{ *} _{j} \}\,\,
~~~~~~( {\rm for}~M_1 \ll M_2, M_3)
 \label{asym} 
\eea
 where $f(x)$ is the loop function eq. (\ref{floop}).

The matrix $\Lambda$ is given by

\beq
\label{k}
\Lambda=\left[
\begin{array}{cc}
\Lambda^{ee} & 0  \\
 0 & \Lambda^{\mu \mu }
\end{array}
\right],
\eeq
where  $\Lambda^{ee}$ and $\Lambda^{\mu \mu}$ indicate
 the {\it complex}  thermal masses
due to the electron and muon Yukawa couplings:

\beq
\Lambda^{ii}\equiv \Lambda^{ii}_\omega -i\, \Lambda^{ii}_{{\rm d}}=
\left. \frac{\omega_{ii}-i\,\Gamma_{ii}}{H(M_1)}\right|_{T=M_1},
\,\,\,i=e,\mu.
\eeq
These will cause the decaying  flavour oscillations
of the asymmetries.
Since the interactions mediated by the electron Yukawa
coupling are out-of-equilibrium, we may safely set $\Lambda^{ee}=0$. On the 
other hand \cite{CKO}

\beq
\omega_{\mu\mu}\simeq \frac{h_\mu^2}{16}\,T,\,\,\,\Gamma_{\mu\mu}
\simeq 5\times 10^{-3} h_\mu^2\,T,
\eeq
leading to

\beq
\Lambda^{\mu\mu}_{{\rm d}}\simeq 3\times 10^{-4}\,h_
\mu^2\,\frac{M_{\rm P}}{M_1},
\eeq
where $M_{\rm P}$ is the Planck mass. Interactions are mediated by the muon Yukawa coupling
 only for $z\ga z_{\rm d}\simeq 
1/\Lambda^{\mu\mu}_{{\rm d}}$. 
Therefore, as thermal leptogenesis takes place at temperatures on the
order of $M_1$, we conclude that -- roughly -- interactions involving
the muon Yukawa coupling
are out-of-equilibrium in the primeval plasma if $M_1\ga 10^{9}$ GeV.

 Following Stodolsky \cite{leo}, we 
may now write eq. (\ref{osc}) in a more compact form 
by expanding the various matrices on the basis provided by the 
 $\sigma$-matrices, $\sigma^\mu=(I,\vec{\sigma})$. A  
generic $2\times 2$ matrix $A$ can be written as
\beq
A= A^\mu \sigma_\mu , ~~~~ A^\mu = \frac{1}{2} 
{\rm Tr} \{A \sigma^\mu\}.
\label{sigmaS}
\eeq
In this basis, the Boltzmann equation (\ref{osc}) becomes a system
of equations of the form 

\begin{eqnarray}
\frac{dY^0_L}{dz}&=&\frac{z}{sH(M_1)}\left( \gamma_D
\left(\frac{Y_{N_{1}}}{Y_{N_{1}}^{eq}} 
- 1\right)\epsilon^0-\frac{1}{Y_L^{eq}}\gamma^0_DY^0_L-\frac{1}{Y_L^{eq}}
\vec{\gamma}_D\cdot\vec{Y}_L\right),\nonumber\\
\frac{d\vec{Y}_L}{dz}&=&\frac{z}{sH(M_1)}\left( \gamma_D
\left( \frac{Y_{N_{1}}}{Y_{N_{1}}^{eq}} -1 \right)
%\frac{Y_{N_{1}}}{Y_{N_{1}}^{eq}} 
\vec{\epsilon}-\vec{\gamma}_D \frac{ Y^0_L}{Y_L^{eq}}
-\frac{1}{Y_L^{eq}}
\gamma^0_D\vec{Y}_L\right)+ \, \vec{\Lambda}_\omega \times \vec{Y}_L
+  \hat{\Lambda}_d \times  \vec{\Lambda}_d \times \vec{Y}_L .
\label{compact}
\end{eqnarray}
Notice, in particular, that $Y^0_L=(1/2) {\rm Tr}\,Y_L$ and therefore
it represents 
 half of the total lepton asymmetry. On the other hand, $Y^z_L=(1/2)
\left(Y_{ee}-Y_{\mu\mu}\right)$
 indicates half of the difference of the asymmetries
in the electron and muon flavour densities. The components $Y_L^x$ and $Y_L^y$
parametrize the off-diagonal entries of the asymmetries and therefore
account for the quantum correlations between the lepton asymmetries.
As 
the matrix $\Lambda$ in eq. (\ref{k})
is diagonal, only the  $\Lambda^0$ and $\Lambda^z$ components are
 different from zero.

A simple inspection of 
eqs. (\ref{compact}) tells us that the components $Y_L^{x}$ and $Y_L^{y}$
precess around the $z$-direction with an angular velocity set by the
thermal mass $\omega_{\mu\mu}$. At the same time, such a precession 
is damped by the interactions mediated by the 
muon-Yukawa coupling at a rate $\sim \Gamma_{\mu\mu}$. 
  Notice that $\omega_{\mu\mu} \sim \Gamma_{\mu\mu}$, so
unlike the case of neutrino oscillations in matter (MSW),
the decoherence and oscillation timescales are the same.
The term
$\vec{\Lambda}\times \vec{Y}_L$ contains all the information about the action
of the decoherent plasma onto the coherence of the flavour oscillations:
if the damping rate $\Gamma_{\mu\mu}$ is much larger than the expansion
rate of the Universe, the quantum correlations among the flavours
asymmetries are quickly damped away. Therefore, we expect that, if 
leptogenesis takes place at  values of $z\gg z_{\rm d}$, quantum
correlations  play no role in the dynamics of leptogenesis \cite{barbieri}.

If thermal leptogenesis occurs  well before 
the muon-Yukawa coupling enters into equilbrium  (that is at $z\ll z_{\rm d}$,
$T\sim M_1\gg 10^{9}$ GeV), then  $|\vec{\Lambda}|z\ll 1$
and the last term of the second equation of (\ref{compact}) can be
safely dropped. In the absence of  the charged lepton Yukawa coupling, 
%eqs. (\ref{compact})  are invariant under flavour rotations and 
we may 
rotate the system in such a way to put all the
asymmetry generated by the decay of the 
right-handed neutrino into a single flavour,
{\it e.g.} $Y^{ee}$. Under these circumstances, $Y^0_L=Y^z_L=(Y^{ee}/2)$,
$Y^x_L=Y^y_L=0$ and similarly for the CP-asymmetries, 
$\epsilon^0=\epsilon^z=(\epsilon^{ee}/2)$,
$\epsilon^x=\epsilon^y=0$. By simply summing the corresponding euqations for
$Y^0_L$ and $Y^z_L$, one immediately finds the Boltzmann equation for the
 single flavour case. 

To get some insight about the dynamics of the system, let 
us now discuss the solution of eqs. (\ref{compact}) in the case in
which thermal leptogenesis takes place nearly in equilibrium which, given the
low-energy neutrino parameters, is likely to be the one realized in Nature.
We parametrise the inverse 
decays with rates $\gamma_D^{ij}=n_N^{eq}\left(K_1(z)/K_2(z)\right)
\Gamma^{ij}$ through the parameters (defined in eq. (\ref{outofeq}))
$K^{ij}\equiv \left(\Gamma^{ij}/H\right)_{T=M_1}$ and work in the limit
in which every $K^{ij}\ga 1$. By defining the combinations
$Y^\pm_L=\left(Y^x_L\pm i Y^y_L\right)/\sqrt{2}$ and by exploiting
 various  saddle-point approximations, we find

\begin{eqnarray}
\label{l}
Y^0_L&\simeq& \frac{\epsilon^0}{g_*}\sqrt{\frac{\pi}{2}}
\frac{1}{K^0\,z_{\rm f}}-\frac{1}{K^0}\left(K^z Y_L^z+K^+Y_L^+ +
K^- Y_L^-\right)(z_{\rm f}),\nonumber\\
Y^z_L&\simeq& \frac{\epsilon^z}{g_*}\sqrt{\frac{\pi}{2}}
\frac{1}{K^0\,z_{\rm f}}-\frac{K^z}{K^0}Y^0_L(z_{\rm f}),\nonumber\\
Y^\pm_L&\simeq& \frac{\epsilon^\pm}{g_*}\sqrt{\frac{\pi}{2}}z_{\rm d}^{3/2}
e^{-z_{\rm d}}{\rm cos}\left[\frac{\Lambda^{\mu\mu}_\omega}{2}\left(
z_{\rm f}^2-z_{\rm d}^2\right)\right]\,e^{-\frac{1}{2}\Lambda_{\rm d}^{\mu\mu}
\left(z_{\rm f}^2-z_{\rm d}^2\right)}\nonumber\\
&-&K^\pm\,Y^0_L(z_{\rm d}) \,z_{\rm d}^{5/2}\,
e^{-z_{\rm d}}{\rm cos}\left[\frac{\Lambda^{\mu\mu}_\omega}{2}\left(
z_{\rm f}^2-z_{\rm d}^2\right)\right]\,e^{-\frac{1}{2}\Lambda_{\rm d}^{\mu\mu}
\left(z_{\rm f}^2-z_{\rm d}^2\right)},
\end{eqnarray}
where $z_{\rm f}\simeq {\rm ln}\,K^0+(5/2){\rm ln}\,{\rm ln}\,K^0$. These
solutions have been obtained in the limit in which $z_{\rm d}\la z_{\rm f}$, 
that is 
when  the muon-Yukawa coupling induces interactions that are in thermal
equilibrium at temperatures larger than the effective temperature at which
inverse decays go out-of-equilibrium. In the limit $z_{\rm d}\ll z_{\rm f}$,
eqs. (\ref{l}) clearly show the decoupling of the quantum correlations from the
dynamics in the limit in which muon Yukawa coupling induced interactions
have come into equilibrium long before thermal leptogenesis, 
$\Lambda^{\mu\mu}z_{\rm f}\gg 1$. In this limit, we find

\beq 
{\rm Tr} Y_L=2\,Y^0_L\simeq \sqrt{\frac{\pi}{2}}\frac{1}{g_*\,z_{\rm f}}
\left(\frac{K^0\epsilon^0-K^z\epsilon^z}{(K^0)^2-(K^z)^2}
\right)=\sqrt{\frac{\pi}{2}}\frac{1}{g_*\,z_{\rm f}}\left(\frac{\epsilon^{11}}{
K^{11}}+\frac{\epsilon^{22}}{
K^{22}}
\right).
\label{asymf}
\end{equation}
As a last comment, let us notice that in the limit $z_{\rm f}\simeq z_{\rm d}$,
the quantum correlations are not efficiently damped out,   and appear
as oscillations in the washout terms of eq.
(\ref{compact}). We expect that the effect 
 on the final total lepton asymmetry,
\beq
Y_L^0\propto \frac{K^0\epsilon^0-\vec{K}\cdot \vec{\epsilon}}{(K^0)^2-
|\vec{K}|^2}~~,
\eeq
will be   analogous to multiplying the washout rate
 by a  neutrino survival probability,  and integrating over time
(an ${\cal O}(1)$ effect).
This case deserves
further investigation and a detailed numerical study \cite{todo}.

\section{Flavour in the Boltzmann Equations}
\label{flavour}
 
In this section,  we consider   Boltzmann equations (BE)  for
the diagonal elements of  $[Y_L]$ with  3 flavours. As discussed in
the previous section  and in  \cite{barbieri}, BE are appropriate
when   the interactions of the charged lepton Yukawa couplings
are much faster, or much slower, than the expansion $H$---provided one
works in the physical basis. To be explicit,
we work  in the charged lepton mass
eigenstate basis, so these equations apply when the $\tau$ and
$\mu$ Yukawa couplings are in equilibrium (see eq. (\ref{yuk})), 
and have caused 
the off-diagonal elements of $[Y_L]$  to be irrelevant. 
 This corresponds to  $T \lsim 10^9$ GeV for the SM (and to $T \lsim 10^9
\times \tan^2 \beta $, where $\tan\beta$ is the ratio of the vacuum expectation
values of the two Higgs bosons, in the supersymmetric version of the SM).

Let us start out by considering  the Boltzmann equation
  as a matrix  equation in  flavour space, 
with only decays and inverse decays, as we did in the previous section,   and   neglecting
\footnote{Including this transformation would
mildly reduce the washout term, because, ${\it e.g.}$,
an $\ell_\mu$ produced is a unit of $B/3 - L_\mu$,
but not all the $B/3 - L_\mu$ is in  $\ell_{\mu}$s. In appendix C we give the explicit tranformation
from $L_i$ to $B/3 -L_i$ for different cases.}
the transformation to $B/3 - L_i$ 
\bea
\frac{sH(M_1)}{z} ~
\frac{d}{dz} 
\left[
\begin{array}{ccc}
Y_{L}^{ee} & & \\
&Y_{L}^{ \mu \mu}
&
\\
& &Y_{L}^{\tau \tau} 
\end{array}
\right]
& = & 
%\frac{z }{s H} 
%\left( 
\gamma_D
\left(\frac{Y_{N_{1}}}{Y_{N_{1}}^{eq}} -1 \right) 
\left[
\begin{array}{ccc}
\epsilon^{e e} & & \\
 &\epsilon^{\mu \mu}
&
\\
& & \epsilon^{\tau \tau}
\end{array}
\right]  %\right.
 \nonumber \\
& & %\left.
 - \frac{1}{2 Y_{L}^{eq}}\left\{ 
\left[
\begin{array}{ccc}
\gamma ^{ee}_D & & \\
&\gamma ^{\mu \mu }_D
&
\\
& & \gamma ^{\tau \tau}_D
\end{array}
\right],
\left[
\begin{array}{ccc}
Y_{L}^{ee} & & \\
&Y_{L}^{ \mu \mu}
&
\\
& &Y_{L}^{\tau \tau} 
\end{array}
\right]
 \right\} %\right)
\label{matrices}
\eea
where $\epsilon^{ii}$ and $\gamma_D^{ii}$ are defined
as in eqs. (\ref{gij}, \ref{asym}).
 Since the off-diagonal elements of the matrices have
been neglected, this equation is not invariant under flavour
rotations, and only applies in the physical basis where
the off-diagonals are suppressed.

We can
illustrate the problem that arises when one sums over  flavour
 by taking the trace of eq. (\ref{matrices}).
The BE for the sum is
\beq
\sum_i \frac{dY^{ii}_{L}}{dz}  =  \frac{z}{s H}\left( \left(
\frac{Y_{N_{1}}}{Y_{N_{1}}^{eq}} - 1\right)
\epsilon \gamma_D
- \sum_i  \gamma^{ii}_D\frac{Y^{ii}_{L}}{Y_{L}^{eq}} \right)
\label{flasum}
\eeq
where $\sum_i \epsilon^{ii} = \epsilon $
is the total CP asymmetry. Like $\gamma_D$  it is the trace
of a matrix in flavour space, see eq. (\ref{gij}).

In the ``single flavour or dominant state approximation''  (see eq. (\ref{YLs})),
the trace of the product of matrices in  the last term on the 
RHS of eq. (\ref{flasum})   is 
approximated as the product of the traces:  
this term  is replaced by $\sum_j Y_{L}^{jj} \sum_i\gamma^{ii}_D$. 
This is correct in the absence of charged lepton Yukawa couplings; if
$\hat{\ell}$ is the flavour combination  into which $N_1$ decays,
then there is a basis where $[\gamma_D] = {\rm diag} (\gamma_D, 0, 0)$
and $[Y_L]_{11} = Y_L$. The trace is invariant under basis transformations 
in flavour space, so for any unitary $V$
\beq
\frac{1}{2} {\rm Tr} \{ V [\gamma_D] V^\dagger,  V [Y_L] V^\dagger \}
 = \gamma_D Y_L ~~~~~~~({\rm no~charged~lepton~Yukawas}).
\eeq
In particular, this is the washout term evaluated in the charged
lepton mass eigenstate basis, see eq. (\ref{YLs}). 

If  the charged lepton
Yukawa couplings are included, then new terms appear in the
equations for the asymmetry \cite{barbieri}, 
as discussed in the previous section.
These new terms suppress the off-diagonals of $[Y_L]$ 
{\it in the charged lepton mass eigenstate basis}, so
the Boltzmann equations (without the off-diagonals) are only appropriate in the
charged lepton mass eigenstate basis,
and in general it is no longer true that
Tr$( [\gamma_D] [Y_L] ) =$ Tr$ [\gamma_D] $ Tr$ [Y_L]$.

One way to see what is being neglected, if the single flavour
approximation is used when the charged lepton Yukawa couplings are in
equilibrium, is to
expand the diagonal matrices $[Y_L], [\epsilon]$ and $[\gamma_D]$
on  the identity matrix $I$  and the diagonal SU(3) generators 
$\Lambda_3 = {\rm Diag} \{ 1, -1, 0 \}$ and $\Lambda_8 = \frac{1}{\sqrt{3}}{\rm Diag} \{ 1, 1, -2 \}$.
For instance, 
\beq
[Y_L] =  \frac{ Y_L}{3} I +  Y_{3} \Lambda_3 + Y_{8} \Lambda_8
\eeq
where  $Y_{3}  =  \frac{1}{2} {\rm Tr} ( Y \Lambda_3 )$ 
 is the flavour difference
$(Y^{ee} - Y^{\mu \mu})/2$. Taking the trace of eq.  (\ref{matrices}), 
we obtain  the BE for the total asymmetry $Y_L$:
\beq
\frac{dY_L}{dz}  =  \frac{z}{s H}\left( 
\left(\frac{Y_{N_{1}}}{Y_{N_{1}}^{eq}} - 1\right)
 \gamma_D \epsilon
- \left(\frac{1}{3} \gamma_D \frac{ Y_{L}} {Y_{L}^{eq}} 
+2 \gamma_{D3} \frac{ Y_{3}} {Y_{L}^{eq}} 
+ 2 \gamma_{D8}\frac{ Y_{8}} {Y_{L}^{eq}} \right)\right).
\label{trace2}
\eeq
In the absence of $\Delta L = 2$ interactions, eq. 
(\ref{matrices}) consists of  three decoupled equations, 
so the final asymmetry can be
obtained easily  by  solving  for each flavour, then
adding the solutions. However, 
we can also obtain Boltzmann
equations for the asymmetries $Y_{3}$  and $Y_{8}$,  by multiplying
eq.  (\ref{matrices}) by $\Lambda_3$ (or  $\Lambda_8$),
and then taking the trace:
\bea
\frac{dY_3}{dz}  & =&  \frac{z}{s H}\left( 
\left( \frac{Y_{N_{1}}}{Y_{N_{1}}^{eq}} -1 \right)
 \gamma_D \epsilon_3
- \left(\frac{ \gamma_D}{3} \frac{ Y_{3}} {Y_{L}^{eq}} 
+ \frac{ \gamma_{D3}}{3} \frac{ Y_{L}} {Y_{L}^{eq}} 
+ \frac{ \gamma_{D3}}{\sqrt{3}}\frac{ Y_{8}} {Y_{L}^{eq}}
+ \frac{\gamma_{D8}}{\sqrt{3}} \frac{ Y_{3}} {Y_{L}^{eq}} 
\right)\right),
 \\
\frac{dY_8}{dz}&  =&  \frac{z}{s H}\left( 
\left(\frac{Y_{N_{1}}}{Y_{N_{1}}^{eq}}-1 \right)
 \gamma_D \epsilon_8 
- \left(\frac{ \gamma_D}{3} \frac{ Y_{8}} {Y_{L}^{eq}} 
+ \frac{ \gamma_{D8}}{3} \frac{ Y_{L}} {Y_{L}^{eq}} 
+ \frac{\gamma_{D3}}{\sqrt{3}} \frac{ Y_{3}} {Y_{L}^{eq}}
- \frac{\gamma_{D8}}{\sqrt{3}} \frac{ Y_{8}} {Y_{L}^{eq}}
\right)\right).
\label{BEasym}
\eea
The factor of $1/3$, multiplying
$\gamma_D Y_L$  in eq. (\ref{trace2}) is interesting. 
One can see from \cite{dip} that in many models,  $N_1$ decays
about equally to all flavours, and produces about equal
asymmetry in all flavours.  One can therefore 
neglect  the terms $\gamma_{D3}  Y_{3} 
+  \gamma_{D8}  Y_{8}$ in  eqs. (\ref{trace2}) and 
 (\ref{BEasym}).
Thus,  leptogenesis is described by the usual equation, with
the  washout  reduced by a factor of 1/3.
This should not be surprising: 1/3 of the asymmetry is
in each flavour, and for each flavour, the (inverse)
decay rate is 1/3 of  the total. Summing over flavour,
$3 \times 1/3 \times 1/3= 1/3$.  We will study
leptogenesis in the context of these flavoured Boltzmann
equations in a subsequent publication \cite{todo}.

\section{Implications of flavours}
\label{applns}

In this section, we  study the modifications
due to flavour
of the leptogenesis bounds on $T_{\rm RH}$ and
the light neutrino mass scale $\bar{m}$.
 In the ``single flavour approximation'', 
an $N_1$-parameter-independent bound 
on the light neutrino mass scale, 
 of $\frac{\bar{m}}{\sqrt{3}} \lsim 0.15$ eV, can be 
obtained \cite{bdp2,HLNPS}.
%{\bf  This bound is
%independent of the $N_1$ parameters.} 
We will
argue below that this bound does not hold  when flavour
effects  are included.
%are relevant, that is when $M_1$ is   smaller than about
%$10^{12}$ GeV, temperature beneath which the interactions mediated
%by the $\tau$-Yukawa coupling enter in equilibrium. In such a range
%of $M_1$, 
The bound 
$m_\nu \lsim  4  \sqrt{10^{10} {\rm GeV}/T_{B - L}}$ eV 
\cite{Fischler:1991gn}, from requiring
$\Delta L = 2$ processes to not wash out the
asymmetry, would apply, but remains a 
function of the (unknown) temperature $T_{B-L}$ at which 
leptogenesis takes place.

The limit $m_\nu \lsim 0.15$ eV, can be
understood  to arise from 
the lower bound on the total decay rate 
$m_1 \leq \tilde{m}_1$,
 and 
the upper bound
on the total CP asymmetry \cite{di,bound}:
\beq
\epsilon \leq   \frac{   3 M_1 \Delta m^2_{atm}} { 8 \pi v^2 m_3},
\label{upper}
\eeq
where $m_3$ is the
largest neutrino mass, and $m_1$ the smallest. 
We assume  the light neutrinos are degenerate,
so $|m_1| \simeq |m_2| \simeq |m_3| \equiv \bar{m}/\sqrt{3}$ --- but
the masses can have different Majorana phases. Leptogenesis
takes place in the strong washout regime, due to
the lower bound on the total decay rate. 
The final baryon asymmetry can be
roughly approximated as
$\eta_B \sim 10^{-3}  \epsilon/K \propto \Delta m^2_{atm}/  \bar{m}^2$.
As the light neutrino mass scale is increased, $M_1$
and the temperature of leptogenesis
must increase 
to compensate the  $\Delta m^2_{atm}/  \bar{m}^2$ 
suppression. However, this
temperature is bounded from above, from the requirement of having
the $\Delta L = 2$
processes  out of equilibrium when
leptogenesis takes place:
\beq
\frac{  \bar{m}^2 T^3}{12 \pi v^4} \lsim \frac{10 T^2}{M_{\rm P}},
\eeq 
so  $M_1 \lsim %T \lsim 100 v^4/( m_\nu^2 m_{pl})  \sim 
10^{10} ({\rm eV}/ \bar{m})^2$ GeV. 
There is therefore an upper bound on the  baryon asymmetry 
which scales as $1/ \bar{m}^4$,  and  with our rough estimates,
one finds $ \bar{m}/\sqrt{3}  \lsim 0.1$ eV.

The individual flavour asymmetries, $\epsilon_{ii}$
do not satisfy the bound of eq. (\ref{upper}).
This can be guessed  from the  ``Jarlskog invariant''
for leptogenesis \cite{DK} \footnote{This
invariant is only applicable when the 
right-handed neutrinos $N_I$ are hierarchical.}, which is the trace:
\bea
I_1 & =& \Im \{ {\rm Tr} [ m_\nu  m_\nu^\dagger (\lambda^T \lambda^*)^{-1}
  m_\nu (\lambda^\dagger \lambda)^{-1}   m_\nu^\dagger ] \}  \nonumber \\
&= &  \Im \left\{  \sum_i  [ U^* D_{m_\nu}^2 W^T D_{\lambda}^{-2} 
W^* D_{  m_\nu} W^\dagger 
D_{\lambda}^{-2}W D_{ m_\nu} U^T ]_{ii} \right\}~~,
\label{LHI1}
\eea
where the unitary matrix $W$ transforms from the
basis where $m_\nu$  is diagonal to
the one where  $\lambda^\dagger \lambda$ is diagonal
($e.g.$ 
in the $D_{\lambda}$ basis, $ m_\nu =
W^* D_{m_\nu} W^\dagger$).
 The total asymmetry
$\epsilon$ is proportional to this trace,
which  vanishes when the light neutrinos are 
exactly degenerate in magnitude. 
However, if  the  invariant is ``cut open'' on lepton
flavour indices, one can show that  the flavour asymmetries 
satisfy 
\beq
|\epsilon_{ii}| \leq \frac{\sqrt{3} M_1 \bar{m}}{8 \pi v^2},
\label{weak}
\eeq
and only the sum vanishes. So 
 if we can arrange  $- \epsilon_{\mu \mu} =  \epsilon_{\tau \tau} \sim
\sqrt{3} M_1 \bar{m}/(8 \pi v^2)$,  and decay rates $\gamma^{\mu \mu}_D = 
s^2  \gamma_{D}$, 
$\gamma^{\tau \tau}_D = c^2  \gamma_{D}$,
then in the  approximation  of strong washout and
neglecting $\Delta L = 2$ processes, 
eq. (\ref{asymf}) gives the total
lepton asymmetry to be
\beq
\label{kk}
Y_L \lsim   \frac{1}{g_*\,z_{\rm f}}
\frac{3 M_1 \tilde{m}_*}{8 \pi v^2} \frac{c^4 - s^4}{c^2s^2}.
\label{big}
\eeq
So we obtain that the baryon asymmetry is independent of the light
neutrino mass scale, under the assumption that
$\Delta L = 2$ processes are negligeable. 
In passing \footnote{We will study the final 
asymmetry as a function of similar angles in a
subsequent publication \cite{todo}.}, one can notice that $Y_L$ increases,
when one of the partial decay widths decreases, provided
that  it remains in the strong washout regime: 
$s^2, c^2 \gg \tilde{m}_*/\tilde{m}$.

We first  show that  flavour asymmetries
are constrained by eq. (\ref{weak}),  when
light neutrinos are degenerate  at mass $\bar{m}/\sqrt{3}$, as opposed
to the stronger  bound on the total asymmetry
of eq. (\ref{upper}).   For hierarchical $N_J$,
with exactly degenerate light neutrinos
\beq
\epsilon_{ii}   =   \frac{3}{8 \pi v^2}
\frac{M_1}{  |\vec{ \lambda}|^2}
 {\rm Im}  \{\lambda_{i} (\vec{\lambda} \cdot  m_\nu^*)_i 
\}   =   \frac{\sqrt{3}M_1\bar{m}}{8 \pi v^2}
\frac{ {\rm Im}  \{\lambda_{i} (\vec{\lambda} \cdot U U^T)_i 
\} }{  |\vec{ \lambda}|^2}
 \leq \frac{ \sqrt{3} M_1\bar{m}}{8 \pi v^2},
\eeq
where there is  no sum on $i$,  
$(\vec{ \lambda})_i = \lambda_{1 i}$,
and  $m_\nu^* = U D_m U^T = \frac{\bar{m}}{\sqrt{3}} U  U^T$.

We can verify that a large neutrino
mass scale is consistent  with leptogenesis
 by explicit construction
of a model. 
It is convenient to use 
the Casas-Ibarra \cite{casasibarra}  parametrisation,
in terms of $m_i$, $M_J$, $U$ and 
the complex orthogonal matrix 
$ R \equiv  v D_M^{-1/2} \lambda U D_m^{-1/2}
$. This ensures we obtain the correct low-energy
parameters.  We take a two-flavour model,
and write $R$ as a rotation through the complex angle 
$\theta = \varphi + i \eta$.
The flavour asymmetries, for maximal atmospheric
mixing and no phase in $U$,  are
\bea
\epsilon_{\mu \mu}&  =& \frac{\sqrt{3} M_1 \bar{m}}{8 \pi v^2}
\frac{1}{\sum_\alpha |\sum_j R_{1j}U^*_{\alpha j}|^2} 
\Im \{ R_{1k} U_{\mu k}^*  R_{1p} U_{\mu p} \}
= -  \frac{\sqrt{3} M_1  \bar{m}}{8 \pi v^2}\frac{\sinh \eta \cosh \eta}{\sinh^2 \eta + 
\cosh^2 \eta} \cos 2 \varphi ~,
 \nonumber \\
\epsilon_{\tau \tau}&  =& \frac{\sqrt{3} M_1  \bar{m}}{8 \pi v^2}
\frac{1}{\sum_\alpha |\sum_j R_{1j}U^*_{\alpha j}|^2} 
\Im \{ R_{1k} U_{\tau k}^*  R_{1p} U_{\tau p} \} 
=   \frac{\sqrt{3} M_1  \bar{m}}{8 \pi v^2}\frac{\sinh \eta \cosh \eta}{\sinh^2 \eta + 
\cosh^2 \eta} \cos 2 \varphi ~.
\label{particular1}
\eea
The flavour dependent decay rates $\gamma_{\mu \mu}$
and $\gamma_{\tau \tau}$ are proportional to
\bea
|\lambda_{1 \mu}|^2 & =& \frac{M_1  \bar{m}}{2v^2\sqrt{3}}| R_{11} + R_{12}|^2  = 
\frac{M_1  \bar{m}}{2v^2\sqrt{3}} (\sinh^2 \eta + 
\cosh^2 \eta +   \sin 2 \varphi) ~,
\nonumber \\
|\lambda_{1 \tau}|^2 & =&  \frac{M_1  \bar{m}}{2v^2\sqrt{3}}| R_{11} - R_{12} |^2 
= \frac{M_1  \bar{m}}{2v^2\sqrt{3}} (\sinh^2 \eta + 
\cosh^2 \eta -  \sin 2 \varphi).
\label{particular2}
\eea
We see that we can choose $\eta$ and $\varphi$ 
as required to obtain eq.  (\ref{big}) with $ c^4-s^4 \sim s^2 c^2$.
We conclude that, if $M_1\la 10^{12}$ GeV and flavours effects are relevant, 
the absolute bound on the light neutrino masses
is the one inferred in ref. \cite{Fischler:1991gn}, 
$m_\nu \lsim  4  \sqrt{10^{10} {\rm GeV}/T_{B - L}}$ eV. Of course, 
if $M_1\ga 10^{12}$ GeV, flavour effects are irrelevant and
we are back to the single flavour case where the
bound $\frac{\bar{m}}{\sqrt{3}} \lsim 0.15$ eV applies\footnote{We thank
 E. Nardi, Y. Nir and E. Roulet for discussions about this point.}.

 The minimum required   reheat  temperature of the Universe,
$T_{\rm RH}$, will  depend  on the flavour structure of the lepton asymmetry.
For thermal leptogenesis to take place,  an
adequate number density of  $N_1$ must be produced by thermal
scattering in the plasma, suggesting  $T_{\rm RH} \gsim  M_1/10$
\cite{bound,leptogen}.
There is a lower bound on $M_1$, from requiring
a large enough CP asymmetry; the corresponding bound on $T_{\rm RH}$ 
depends on the fine details of reheating, whose dynamics 
have been addressed in
ref. \cite{leptogen}.
The lower bound  of \cite{leptogen},
$T_{\rm RH} \gsim (2-3) \times 10^{9}$ GeV,
was obtained with hierarchical light neutrinos
and strong washout. We include flavour effects by
rescaling this bound.

In many popular seesaw models, the
light neutrinos are hierarchical, 
leptogenesis takes place in the strong
washout regime $ K > 1$, and the
asymmetries and decay rates are similar
for all flavours: $\gamma_D^{ii} \sim \gamma_D^{jj}$
and $\epsilon^{ii} \sim \epsilon^{jj}$.  In this
case, $\epsilon$ is bounded by (\ref{upper}), but
the washout is reduced by the factor of 1/3
from equation (\ref{trace2}), so we estimate
that
\beq
T_{\rm RH} \gsim 6 \times  10^{8} {\rm GeV} ~~.
\eeq

Thermal leptogenesis could work with a lower
reheat temperature 
%in the case of degenerate light neutrinos, 
if the washout rate in one flavour is suppressed. 
If $K > 1$, a thermal distribution of $N_1$s
will be created, but the asymmetry in flavour $j$
is only washed out by $\gamma_D^{jj}$, so
\footnote{This formula is valid in the strong washout 
regime, so requires $K^{jj} = \gamma^{jj}K/\gamma_D >1$.}
$
Y_L^{jj} \sim 10^{-3} \epsilon^{jj}/K^{jj}$.
For instance, in our ``degenerate'' two generation model,
we can tune the couplings \footnote{For
complex $U_{\mu 3} = U_{\tau 3} = e^{i \delta}/\sqrt{2}$,
  with $\phi$ very close to $-\pi/4$
and $\sinh\eta = -\cosh\eta \sin\delta/(1+\cos\delta) $we obtain small
 $\lambda_{1 \mu}$. 
}  so that
$\lambda_{1 \mu}/\lambda_{1 \tau}  \simeq s $ is
of order $\sqrt{\tilde{m}_*/\bar{m}}$ where 
$\bar{m} \sim$ eV, in  a generous
intepretation of the cosmological bound \cite{cosmo}. This suppresses
$\epsilon^{\mu \mu}$ with respect to
eq. (\ref{big}) by  $s$, so
the  final asymmetry  $Y_L \sim Y_L^{\mu \mu}$ scales
as $1/s$ and  we obtain
\beq
T_{\rm RH} \gsim  2 s \times 10^{9} {\rm GeV}  \simeq 
10^{8} {\rm ~GeV ~~~for~~tuned,~~ degenerate~~ light ~~ neutrinos}.
\eeq
This  appears theoretically unmotivated,
compared to other    leptogenesis scenarios
at low  $T_{\rm RH}$ \cite{PU,marta}.

Our results can therefore relax
the tension between thermal leptogenesis and the gravitino bound
\cite{gravitino} in supersymmetric extensions of the SM which
impose an upper  bound on $T_{\rm RH}$ of about $10^{9}$ GeV.

\section{Summary}

Including flavour in the analysis of leptogenesis modifies the equations
of motion (Boltzmann Equations)  for the asymmetry, and 
changes some of the relations between parameters
of the calculation. 

The interactions related to the charged Yukawa couplings 
may induce quantum oscillations amongst the lepton asymmetries. 
This interesting result, in combination with the temperature 
dependent relation between chemical potentials,  makes ${\cal O}(1)$ 
changes to   the final value of the baryon asymmetry. 
At temperatures such that  the quantum oscillations 
can be neglected,  we  find that the usual washout
terms in the Boltzmann Equations are modified, and that there are 
additional terms proportional to flavour asymmetries. 

The  upper bound on the CP asymmetry  produced for each flavour 
is weaker than the bound applicable to the total CP asymmetry.
In particular, it increases with the neutrino mass scale. 
By judicious choice of flavour-dependent washout rates,
one can obtain a final baryon asymmetry which is
independent of the light neutrino mass scale.
Finally, given the experimental value of the Baryon Asymmetry of the 
Universe, we find a reduced  lower bound on the lightest right-handed neutrino
mass and on the reheating temperature.

\subsection*{Acknowledgements}

We are happy to thank the authors of ref. \cite{nnrr} for communicating their
results to us prior to publication. We 
are grateful to  G. Giudice, A. Ibarra, 
E. Nardi, Y. Nir, A. Pilaftsis, M. Pl\"umacher, E. Roulet, 
A. Strumia, 
T Underwood and O. Vives  for useful comments and 
discussions. 
AR is on leave of absence from INFN, Padova.

\appendix

\section{ From a toy model to equations 
of motion for $f_\Delta^{ij}$}
\label{apenrho}

The aim of this appendix is to obtain 
equations of motion for a system of
quantum  harmonic oscillators, 
which carry  two ``flavours''.   The 
oscillators are  coupled to match the 
interactions of the Lagrangian in eq. (\ref{L}).
By taking expectation values of the number
operator in a ``thermal'' state,  
this model  can be reduced to  a two-state
quantum system. We  then extrapolate
 the flavour structure
of these equations of motion, to 
obtain ``flavoured'' equations of motion
for the lepton asymmetry number operator
in the early Universe.  We will  discuss the
field theory derivation of such equations
in a later publication \cite{todo}.

Consider  a flavour-dependent number operator,
which we call $\hat{f}$,
for a system consisting of two simple harmonic
oscillators labelled  $e$ and $\mu$:
\beq
\hat{f}  = \left[  \begin{array}{cc}
a^\dagger_e a_e & a^\dagger_e a_\mu \\
a^\dagger_\mu a_e & a^\dagger_\mu a_\mu 
\end{array}
 \right],\\\\
\eeq
where the commutation\footnote{In studying flavour, we consider
``bosonic'' leptons. The generalization to the 
realistic case of fermionic leptons
is straightforward.}   relations
  are $[a_e,a_e^\dagger] = 1$, $[a_\mu,a_\mu^\dagger] = 1$,
and the hamiltonian is $H_0 = (\omega_e a^\dagger_e a_e
+\omega_\mu a^\dagger_\mu a_\mu) I $.
The number operator evolves according to the Hamiltonian equations of motion:
$\frac{d \hat{f} }{dt}$ $ =  + i [H,\hat{f} ] $.
Since our Hamiltonian conserves particle number, we can
take  expectation values---for instance,
in a thermal bath---and reduce our toy model to a two
state system, described by $f   \equiv \langle \hat{f}  \rangle$,
satisfying
\bea
\frac{d f   }{dt} 
  & = &  - i \left[ \begin{array}{ccc}
 0 & (\omega_e - \omega_\mu)  f ^{e \mu } 
  \\
 (\omega_\mu - \omega_e)  f ^{\mu e}  & 0
\end{array} \right].
\label{dry}
\eea
Notice that quantum mechanically,
$f $ is  the density matrix of the two-state system.
If at $t= 0$  the system is created
in some  state, then
the probability to be found at a time  $t$ later  in this state
 is $  {\rm Tr} \{  f (0)  f  (t) \}$, which can
oscillate.

The excitations
of the oscillators can be
imagined to be the lepton {\it asymmetry}, carried by
particles  of  energies $\omega_\mu  \sim \omega_e \sim E$. 
Then $\omega_\mu  - \omega_e \simeq m_\mu^2(T)/E$,
where   $m^2_\mu(T) = h_\mu^2 T^2/16 $ is the contribution of
$h_\mu$ to the ``thermal mass''
of $\ell_\mu$ in the early Universe.
Suppose the production  and washout of
the lepton asymmetry  are treated as initial condition and
subsequent measurement on the system. That is,
we imagine to  produce the asymmetry
in some linear combination of $e$ and $\mu$,
allow it to evolve, and then at a later
time turn on the inverse decays. 
Then  the inverse decay rate could be
reduced (or increased)
because  the asymmetry  changed flavour during time evolution
(analogous to a survival probability in neutrino oscillations).

We have neglected the thermal masses $\propto \lambda_i \lambda^*_j$
because we anticipate that its effects are negligeable
at temperatures $\sim 10^9$ GeV: if $|h_\mu|^2 \gg
|\vec{\lambda}|^2$, the $\lambda$-contribution to $H_0$ is small
and can be ignored. If $|h_\mu|^2 \ll
|\vec{\lambda}|^2$, then the asymmetry is produced in
the flavour direction $\hat{\lambda}$, so a thermal mass for
this direction will not cause oscillations (asymmetry in a mass
eigenstate). For a detailed discussion of ``neutrino flavour
effects''  (due  to $\lambda$) in resonant leptogenesis,
see \cite{PU}.

The decays and inverse decays due to the Yukawa interactions  can be
included by perturbing in an  interaction Hamiltonian, $H_I$. 
We first consider  the production and washout of
leptons, due to  $\lambda$, in a  model of
  four harmonic oscillators: $N,H,$ and two leptons flavours 
$\{ \ell_i \}$. Later we will
treat the $N$ and $H$ number densities as
background fields, and take the
difference with respect to the equation
for anti-lepton flavours.  We are looking
for behaviours analogous to  the 
lepton number production and washout in
the decay/inverse decay  $N \rightarrow H \ell$.

We suppose the $H$ and $\ell$ are
massless and $N$ is massive, and ignore the  free Hamiltonian
(considered previously)
which does not change the particle numbers.
The  interaction Hamiltonian 
is 
\bea
H_{I}  & =&
\lambda_k a_N^\dagger  a_H   a_\ell^k  + \lambda^*_k a_N a_H^\dagger  
  a_\ell^{k \dagger}.
\label{HIY}
\eea
We perturbatively expand the Heiseberg
equations of motion
to get that
\bea
\frac{\partial }{\partial t} f_\ell^{ij} 
= -  [ H_{I}, [H_{I} ,f_\ell^{ij} ]] &=& - \lambda^*_k \lambda_i  
[a_N a_H^\dagger a_\ell^{k \dagger}, a_N^\dagger
    a_H   a_\ell^j]  
 +
 \lambda_k \lambda^*_j   [a_N^\dagger  a_H a_\ell^k  , 
a_N   a_H^\dagger  a_\ell^{i \dagger}]  \nonumber\\
&= & 
 -\lambda_i  \lambda^*_k ( f_H f_\ell^{k j} -
 f_N f_H \delta ^{ k j}
- f_N f_\ell^{j k} - f_N \delta ^{j k})
\nonumber \\ && +
 (  f_N \delta ^{ik}
+  f_N f_H \delta ^{ik}
+ f_N f_\ell^{ik}
- f_H f_\ell^{ ik} ) \lambda_k \lambda^*_j ~. 
\label{longdamp}
\eea
This looks like inverse decays and decays of $N$, 
and the middle two terms of each parentheses
 we drop because they
  would not be allowed kinematically for $N$ decay in the
early Universe.
A similar calculation \footnote{We neglect possible
interference terms between $H_{I} $ and
$H_{I, h}$, because the expectation value
of terms such as $a^\dagger_{\mu^c} a_N$ will vanish.}  can be performed
for the  muon Yukawa coupling $h$, with
\bea
H_{I,h}  & =&
h_k a_H^\dagger  a_{\mu^c}   a_\ell^k  + h^*_k a_H 
a_{\mu^c}^\dagger  
  a_\ell^{k \dagger},
\eea
and using the 
kinematic constraint that 
 the processes $H \leftrightarrow \ell_\mu  \mu^c$
 are  allowed, because the thermal mass of Higgs
has contributions from the top.
Taking expectation values, this gives in the
charged lepton mass eigenstate basis:
\bea
\frac{\partial }{\partial t} \left[
\begin{array}{cc}
f_\ell^{ee} & f_\ell^{e \mu} \\ 
f_\ell^{\mu e} &  f_\ell^{\mu \mu}
\end{array} \right] & =&   -  f_{\mu^c} 
\left[
\begin{array}{cc}
0 & 0 \\ 
0 &  |h_\mu|^2
\end{array} \right]
\left[
\begin{array}{cc}
f_\ell^{ee} & f_\ell^{e \mu} \\ 
f_\ell^{\mu e} &  f_\ell^{\mu \mu}
\end{array} \right]
 - f_{\mu^c}\left[
\begin{array}{cc}
f_\ell^{ee} & f_\ell^{e \mu} \\ 
f_\ell^{\mu e} &  f_\ell^{\mu \mu}
\end{array} \right]
\left[
\begin{array}{cc}
0 & 0 \\ 
0 &  |h_\mu|^2
\end{array} \right] \nonumber \\ & & 
 + 2 f_H  \left[
\begin{array}{cc}
0 & 0 \\ 
0 &  |h_\mu|^2
\end{array} \right].
\label{decohere}
\eea
It is reassuring that 
the $t$-independent equilibrium solution 
of this equation  imposes
 $f_{\mu^c}f_\ell^{\mu \mu} - f_H = 0$.
Combined with eq. (\ref{dry}), this gives
damped oscillations for the off-diagonals:
$$ \partial f^{e \mu}_\ell/\partial t = -i ( \omega_e - \omega_\mu -
i |h_\mu|^2f_{\mu^c} ) f^{e \mu}_\ell,
$$ so we can extrapolate that
exchanging doublet with  singlet leptons destroys
the coherence between different doublet flavours.

Equations for the anti-lepton density
can similarly be derived.  For  simplicity,
we take $H$ and $N$ to be their own
anti-particles, so  there are now
eight coupled oscillators: $N, H, \ell^{i}, \bar{\ell}^i
\mu^c$ and $\bar{\mu}^c$. 
The expectation
value of the  difference between the two equations,
again in the charged lepton mass eigenstate basis, is

\bea
\frac{\partial }{\partial t} f_{\Delta }^{ij} 
&=&  -  \lambda_i  \lambda^*_k f_H f_{\Delta }^{k j} 
 - f_H f_{\Delta }^{ i k}  \lambda_k \lambda^*_j
 + 2 (f_N - f_N^{eq}) \epsilon^{ij}
 \nonumber \\
&& - i (\omega_i - \omega_j) f_{\Delta }^{ij}
 - \frac{|h_\mu|^2}{2} (f_{\mu^c}
+ f_{\bar{\mu}^c}) ( f_{\Delta }^{ i \mu}  \delta_{\mu j}
 + \delta_{\mu i}   f_{\Delta }^{\mu j} 
)
 \nonumber \\
&&
- \frac{|h_\mu|^2}{2} (f_{\mu^c}
- f_{\bar{\mu}^c}) (( f_{\ell } + f_{\bar{\ell} }^T) ^{ i \mu}  
 \delta_{\mu j} 
+ \delta_{\mu i}  ( f_{\ell } + f_{\bar{\ell} }^T)^{\mu j} 
),
\label{damp}
\eea
where $k$, $i$, $j$ can be $e$ or $\mu$,
 and  $i$,  $j$ are not summed.
The purpose of this equation is to motivate the flavour index structure
we  use
in the text. 
We now briefly discuss the various terms.
\begin{itemize}
\item The first two terms describe washout by inverse decays.
By construction, there is no washout by $\Delta L = 2$ scattering
(we will include this by hand).
\item The third  term, which 
is  $ \propto \epsilon$,
has been obtained artificially.
It is well-known  in field theory that 
no asymmetry is  created at second order in $\lambda$; % creation is
%proportional to 
the difference $\Gamma (N \rightarrow H \ell)
 - \Gamma (N \rightarrow \bar{H}  \bar{\ell})$
is proportional to 
the imaginary parts of the  loop amplitude times the loop
coupling constants.
 We write the latter as $(\delta \lambda)_i$,
replace  the ``tree-level'' $\lambda_i$ by
a ``tree + loop'' vertex  $ 
\lambda'_i = \lambda_i + (\delta \lambda)_i$,
so
 $ \Gamma - \bar{\Gamma}$
becomes  the matrix in flavour space $
[ (\delta \lambda)_i \lambda^*_j -$ $ 
\lambda_i (\delta \lambda)^*_j]$.
This gives 
\beq
\epsilon_{ ij} = \Im \{  \lambda_{i} (\delta \lambda)^*_{j} -
 (\delta \lambda)_{i} \lambda_{j}^* \}
=\Im \{
\lambda_{i}   \kappa^*_{ jk} 
\lambda_k 
- {\kappa}_{ik} \lambda^{* k}  \lambda^{ *} _{j}\}
\label{guess1}
\eeq
where $[\kappa]_{ij} = [m_\nu]_{ij}/v^2$.
Using the equilibrium condition
$$
2  f_N^{eq} [\epsilon] =  
\frac{ f_H}{2} ( [\epsilon] [f_{\ell} + f_{\bar{\ell}}^T]
+
 [f_{\ell} + f_{\bar{\ell}}^T ][\epsilon]),
$$ (where square brackets are matrices in flavour space),
we obtain a production term
$ 2 (f_N  + f_N^{eq})  \epsilon^{ij}$.
This is incomplete,  because it 
 would produce an asymmetry in thermal
equilibrium.  The resonant part of $\Delta L = 2$
scattering should be subtracted \cite{KW}, which ensures that
Tr $f_\Delta$ vanishes in thermal equilibrium. We
work in the charged lepton mass eigenstate basis, 
and follow \cite{KW} to obtain  eq. (\ref{damp}).
\item The remaining terms of eq. (\ref{damp}) describe the effects of
the muon Yukawa coupling.   The asymmetry  oscillates in
flavour space, as described at the beginning of the section,
because the off-diagonal flavour matrix elements have
time-dependent phases. 
\item The muon Yukawa coupling also
causes the off-diagonal elements to decay away, via the second to last term.
To get the correct relative normalisation between oscillation
and decay in the early Universe,  we will extract the rates from
the  real and imagainary parts of the thermal 
propagator. 
\item The last term has
minimal effect on the off-diagonal terms, as one can
see by solving eq. (\ref{decohere}). However, combined
with the second to last term, it drives
the diagonal asymmetry $Y_{\Delta \ell_\mu}$ to the correct  relation
with  $Y_{B/3 - L_\mu}$,  when the muon Yukawa coupling is
in equilibrium \footnote{In our toy model, where
$H$ is a real scalar, this means $  (f_{\mu^c}
- f_{\bar{\mu}^c})  =  f_{\Delta }^{ \mu \mu}$.
In the SM, there is  Higgs asymmetry \cite{HT, barbieri}.}.
 For simplicity, we drop this term,
assuming  its effects could be approximated
 as time-dependent components
in the $A$-matrix \cite{barbieri} relating
 $  Y_{\Delta \ell_j} =  A^{ji} Y_{B/3 - L_i}$ :
\beq
A^{ij}(t) = A^{ij}_{T>}e^{- \Gamma_\mu t} +   A^{ij}_{T<}(1 - e^{- \Gamma_\mu t}),
\eeq 
where $A^{ij}_{T>}$ are the matrix elements before $h_\mu$ comes
into equilibrium,  $A^{ji}_{T<}$ are the matrix elements afterwards, and
$\Gamma_\mu$ is the interaction rate associated to the muon
Yukawa coupling, see also appendix C. 
\end{itemize}

\subsection*{Added Discussion}

The effect of  the charged lepton Yukawa in equation (\ref{damp}) 
can be  clarified by  studying the evolution of   $ F^{ij} =  f_\Delta^{ij} - 
\frac{h_i h_j^*}{|h|^2}(f_{\mu^c} - f_{\overline{\mu}^c})$, where
$h_i$ is the muon Yukawa coupling in some arbitrary  basis
for lepton doublet space. The trace of $F$ is   the total  lepton
number stored in $e_L$, $\mu_L$ and $\mu_R$, so studying
$F$ in the toy model is analogous to replacing  the asymmetry
in  lepton doublets  with the asymmetry in $B/3 - L_\alpha$
\cite{barbieri} in  usual leptogenesis calculations.   Neglecting
the  first four terms of eqn (\ref{damp}), which describe
  the interactions of
$N$  and the oscillations due to the muon Yukawa,  the
last two terms of eqn (\ref{damp}) can be written 
\bea
\frac{\partial }{\partial t} [f_\Delta] 
&= &
-  
 {\Big \{ } [\vec{h} \vec{h}^\dagger], [f^\Delta]  {\Big \} } 
 \frac{(f_{\mu^c} + f_{\overline{\mu}^c} )}{2}
-   {\Big \{ } [\vec{h} \vec{h}^\dagger], [f_\ell + {f}_{\overline{\ell}}^T]  
{\Big \} }  \frac{(f_{\mu^c} - f_{\overline{\mu}^c} )}{2} ~. 
\eea
where $\vec{h}$ is a colomn vector, and objects in square brackets
are matrices in doublet space. 
The rate of   change of
the $\mu^c$ asymmetry $f_{\Delta \mu^c} = f_{\mu^c} - f_{\overline{\mu}^c}$
 is
\bea
\frac{\partial }{\partial t}f_{\Delta \mu^c} 
 &=&  - |h|^2 (f_{\mu^c}
+ f_{\bar{\mu}^c})  f_{\Delta }^{ \mu \mu}   
- |h|^2  f_{\Delta \mu^c}  ( f_{\ell } + f_{\bar{\ell} }^T) ^{  \mu \mu}  
 \nonumber 
\eea
So the equation for the lepton  asymmetry 
$F^{ij}$, in the flavour basis, is
\bea
\frac{\partial }{\partial t} [F] 
&= &
-  
 \left( {\Big \{ } [\vec{h} \vec{h}^\dagger], [f^\Delta]  {\Big \} }  - 
2 [\vec{h} \vec{h}^\dagger]  \hat{h}^\dagger 
\cdot[f^\Delta]\cdot \hat{h} \right)
 \frac{(f_{\mu^c} + f_{\overline{\mu}^c} )}{2}
\nonumber \\ &&-  
\left( 
 {\Big \{ } [\vec{h} \vec{h}^\dagger], [f_\ell + {f}_{\overline{\ell}}^T]  
 {\Big \} }
-  2 [\vec{h} \vec{h}^\dagger] \hat{h}^\dagger \cdot
[ ( f_{\ell } + f_{\bar{\ell} }^T)]\cdot \hat{h} \right)
  \frac{(f_{\mu^c} - f_{\overline{\mu}^c} )}{2} ~. 
\label{50}
\eea
where the second line is of the same order as the first (see
eqn  (\ref{damp})).%  As discussed in
%the last point after eqn  (\ref{damp}), an equation
%for $F^{jj}$ can be obtained by 
% approximating
%$f_{\Delta \mu^c}$ and $f_\Delta^{jj}$ in terms of  $F^{jj}$
%using a time-dependent A-matrix. 
 As discussed in
the last point after eqn  (\ref{damp}), an approximate 
equation for $F^{jj}$ can be obtained by 
expressing $f_{\Delta \mu^c}$ and $f_\Delta^{jj}$ in terms of  $F^{jj}$
using a time-dependent A-matrix. To obtain
equation  (\ref{osc}), we simply drop the second line. 
In the flavour basis, eqn (\ref{50}) can be written as
\bea
\frac{\partial }{\partial t} \left[
\begin{array}{cc}
f_\Delta^{ee} & f_\Delta^{e \mu} \\ 
f_\Delta^{\mu e} &  f_\Delta^{\mu \mu} -  f_{\Delta \mu^c} 
\end{array} \right] & =&   
- \frac{ |h_\mu|^2}{2 } \left[
\begin{array}{cc}
0 &   f_\Delta^{e \mu} \\ 
 f_\Delta^{\mu e} &  0   
\end{array} \right]
(f_{\mu^c} + f_{\overline{\mu}^c} )
% \nonumber \\ && 
- \frac{ |h_\mu|^2}{2 }  \left[
\begin{array}{cc}
0  &  f_\ell^{e \mu}
+  f_{\overline{\ell}}^{ \mu e} \\ 
 f_\ell^{ \mu e}+  f_{\overline{\ell}}^{ e \mu}
 &   0 
\end{array} \right]
 f_{\Delta \mu^c} \nonumber
\eea
As expected, the Yukawa coupling $h$ causes the
flavour off-diagonal  asymmetries 
to decay away, as it did for  the  off-diagonal number densities
in eqn (\ref{decohere}).  It has no effect
on the asymmetry in either flavour. 
If  the matrices $F$ and   $[\vec{h} \vec{h}^\dagger] = H_a \sigma^a$
are expanded on Pauli matrices, as in equation (\ref{sigmaS}), 
this  behaviour can be modelled as  a triple cross product: 
$\hat{H} \times \vec{H} \times \vec{F}$. A single cross
product, as in the orginal version of the paper, obviously
does not cause the off-diagonals to be damped as
$\partial F^\perp/\partial t  \propto |h|^2  F^\perp$.

\section{ $\Delta L=2$ terms}

Following the same procedure of section \ref{flavour} we can include the contributions
arising from the $\Delta L =2$ exchange mediated by RH neutrinos.
 In matrix form, the additional terms on the RHS of eq. 
(\ref{matrices}) arising from the $s$-channel contribution would be

\bea
{\rm RHS}\,\,{\rm of} \,\,{\rm eq.}\,\, (\ref{matrices})
& =&
  \frac{1}{2 Y_{L}^{eq}}\left\{ 
\left[
\begin{array}{ccc}
\gamma ^{eeee}_{\Delta L=2} & & \\
&\gamma ^{\mu \mu e e }_{\Delta L=2}
&
\\
& & \gamma ^{\tau \tau ee}_{\Delta L=2}
\end{array}
\right],
\left[
\begin{array}{ccc}
Y_{L}^{ee} & & \\
&Y_{L}^{ \mu \mu}
&
\\
& &Y_{L}^{\tau \tau} 
\end{array}
\right]
 \right\} \nonumber \\
& & + \frac{1}{2 Y_{L}^{eq}}\left\{ 
\left[
\begin{array}{ccc}
\gamma ^{ee\mu\mu}_{\Delta L=2} & & \\
&\gamma ^{\mu \mu \mu\mu }_{\Delta L=2}
&
\\
& & \gamma ^{\tau \tau \mu\mu}_{\Delta L=2}
\end{array}
\right],
\left[
\begin{array}{ccc}
Y_{L}^{ee} & & \\
&Y_{L}^{ \mu \mu}
&
\\
& &Y_{L}^{\tau \tau} 
\end{array}
\right]
 \right\} \nonumber \\
& & +\frac{1}{2 Y_{L}^{eq}}\left\{ 
\left[
\begin{array}{ccc}
\gamma ^{ee\tau\tau}_{\Delta L=2} & & \\
&\gamma ^{\mu \mu \tau \tau }_{\Delta L=2}
&
\\
& & \gamma ^{\tau \tau \tau\tau}_{\Delta L=2}
\end{array}
\right],
\left[
\begin{array}{ccc}
Y_{L}^{ee} & & \\
&Y_{L}^{ \mu \mu}
&
\\
& &Y_{L}^{\tau \tau} 
\end{array}
\right]
 \right\} \nonumber \\
& & +\frac{1}{ Y_{L}^{eq}}\left[ \begin{array}{ccc}
{\rm Tr} [\gamma^{eejj}_{\Delta L=2}Y_{L}^{jj}] & & \\
& {\rm Tr} [\gamma^{\mu \mu jj}_{\Delta L=2} Y_{L}^{jj}] & \\
& & {\rm Tr} [\gamma^{\tau\tau jj}_{\Delta L=2} Y_{L}^{jj}]
\end{array}\right],
\label{matricescont}
\eea

where 

\beq
\gamma^{iijj}_{\Delta L=2} = \sum_J \gamma^{iijj J}_{\Delta L=2} = \sum_J |\lambda_{Ji}|^2 |\lambda_{Jj}|^2 F_{JJ}.
\eeq
The sum is over the RH exchanged neutrino eigenstates and the function $F_{JJ}$ is the kinematical factor corrected according to \cite{leptogen}. We define the matrices 

\bea
[\gamma^i_{\Delta L=2}] = \left[\begin{array}{ccc}
\gamma^{iiee}_{\Delta L=2} & & \\
& \gamma^{ii\mu\mu}_{\Delta L=2} & \\
& & \gamma^{ii\tau\tau}_{\Delta L=2}
\end{array} \right],
\eea

and

\bea
[\gamma^{,j}_{\Delta L=2}] = \left[\begin{array}{ccc}
\gamma^{eejj}_{\Delta L=2} & & \\
& \gamma^{\mu\mu jj}_{\Delta L=2} & \\
& & \gamma^{\tau\tau jj}_{\Delta L=2}
\end{array} \right].
\eea
Each of which can be decomposed in terms of the matrices $I,\Lambda_3, \Lambda_8$:

\beq
[\gamma^{i}_{\Delta L=2}] = \frac{1}{3}\gamma^{i}_{\Delta L=2} I + \gamma{i}_{\Delta L=2,3} \Lambda_3 + \gamma^{i}_{\Delta L=2,8} \Lambda_8
\eeq
and similarly for $\gamma^{,j}_{\Delta L=2}$.
The interference terms can also be dealt with in terms of the quantities

\beq
\gamma^{iijj (IJ)}_{\Delta L=2} = \sum_{I<J} [\lambda_{Ii}\lambda^{*}_{Ij}][\lambda^{*}_{Ji}\lambda_{Jj}] G_{IJ}, 
\eeq
where $G_{IJ}$ is a kinematical factor of the same kind as $F_{IJ}$~ \cite{leptogen}.
Thus taking the trace the contribution to eq. (\ref{trace2})

\bea
{\rm RHS} &=& \frac{1}{3}\left[ \gamma^{e}_{\Delta L=2} + \gamma^{\mu}_{\Delta L=2} + \gamma^{\tau}_{\Delta L=2} + \gamma^{,e}_{\Delta L=2} + \gamma^{,\mu}_{\Delta L=2} + \gamma^{,\tau}_{\Delta L=2}\right] Y_{L}
 \nonumber \\
& & + \frac{1}{3}\left[ \gamma^{e (IJ)}_{\Delta L=2} + \gamma^{\mu (IJ)}_{\Delta L=2} + \gamma^{\tau (IJ)}_{\Delta L=2} + \gamma^{,e (IJ)}_{\Delta L=2} + \gamma^{,\mu (IJ)}_{\Delta L=2} + \gamma^{,\tau (IJ)}_{\Delta L=2}\right] Y_{L} \nonumber \\
& & + 2\left[
\gamma^{e}_{\Delta L=2,3} + \gamma^{\mu}_{\Delta L=2,3} + \gamma^{\tau}_{\Delta L=2,3} + \gamma^{,e}_{\Delta L=2,3} + \gamma^{,\mu}_{\Delta L=2,3} + \gamma^{,\tau}_{\Delta L=2,3}\right] Y_{3} \nonumber \\
& & + 2\left[ \gamma^{e (IJ)}_{\Delta L=2,3} + \gamma^{\mu (IJ)}_{\Delta L=2,3} + \gamma^{\tau (IJ)}_{\Delta L=2,3} + \gamma^{,e (IJ)}_{\Delta L=2,3} + \gamma^{,\mu (IJ)}_{\Delta L=2,3} + \gamma^{,\tau (IJ)}_{\Delta L=2,3}\right] Y_{3}\nonumber \\
& & + 2\left[ \gamma^{e}_{\Delta L=2,8} + \gamma^{\mu}_{\Delta L=2,8} + \gamma^{\tau}_{\Delta L=2,8} + \gamma^{,e}_{\Delta L=2,8} + \gamma^{,\mu}_{\Delta L=2,8} + \gamma^{,\tau}_{\Delta L=2,8}\right] Y_{8} \nonumber \\
& & +2 \left[ \gamma^{e (IJ)}_{\Delta L=2,8} + \gamma^{\mu (IJ)}_{\Delta L=2,8} + \gamma^{\tau (IJ)}_{\Delta L=2,8} + \gamma^{,e (IJ)}_{\Delta L=2,8} + \gamma^{,\mu (IJ)}_{\Delta L=2,8} + \gamma^{,\tau (IJ)}_{\Delta L=2,8}\right]  Y_{8} ~~.
\eea
We next show that for low temperatures ($z\gg 1$), the $\Delta L=2$ terms can be neglected
with respect to the inverse decays. To simplify the discussion we focus on  the ID and
$\Delta L =2$ contributions to a single flavour, say elements {11} of the correspondings matrices
that enter the BE. For $z\gg 1$

\beq
\gamma_D^{ee} Y_L^{ee}\sim \frac{e^{-z}}{\pi^{5/2} z^{3/2}} M_1^4 \lambda_{1e} \lambda_{1e}^{*} Y_L^{ee},
\eeq
and the $\Delta L=2$ terms  give,

\beq
{\rm RHS}\,\,{\rm of} \,\,{\rm Eq.} (\ref{matricescont})  \propto \frac{1}{\pi^5 z^6} M_1^4 \left( 2|\lambda_{1e}|^2| \left(\sum_k \lambda_{1k}|^2\right) Y_L^{ee} 
+  |\lambda_{1e}|^2 |\lambda_{1\mu}|^2| Y_L^{\mu\mu} +  |\lambda_{1e}|^2|\lambda_{1\tau}|^2| Y_L^{\tau\tau}\right).
\eeq

This means that for temperatures $z \la z_{\rm c}$,  we can neglect the contribution to the BE from the lepton number violating exchange
of the RH neutrinos as the comparative strength of the inverse decays suppression is $  \sim e^{-z}$ while
the suppression from the $\Delta L=2$ is

\beq
\sim \frac{1}{\pi^{5/2} z^{9/2}} \left(2 \sum_k |\lambda_{1k}|^2 +|\lambda_{1\mu}|^2 \frac{Y_{L}^{\mu\mu}}{Y_L^{ee}} + |\lambda_{1\tau}|^2 \frac{Y_L^{\tau\tau}}{Y_L^{ee}}\right).
\eeq

This implies that for $\sum_{k} |\lambda_{1k}|^2 \sim 1$ only for $z \ga 15$ will the $\Delta L=2$ terms dominate
over $\gamma_D$. For smaller values of $\lambda_{1k}$, as constrained by the departure from thermal equilibrium and
neutrino masses, the $\Delta L=2$ terms are always neglible for the more interesting values of $z$. That is,
 if the freeze out temperature is denoted by $z_{\rm f}$, then 
for the regions in
 parameter space for which $z_{\rm f} \la z_{\rm c}$ the produced asymmetry is not affected by the $\Delta L=2$ washout
terms and thus can be neglected.

\section{ Chemical Equilibrium}

An important point which was pointed out in ref. \cite{barbieri} 
is the relationship between
the lepton asymmetry for each flavour $L_{i}$ and the $B/3-L_i \equiv \Delta_i$ asymmetry, which is what is effectively conserved by the sphaleron interactions, from chemical equilibrium equations.
More recently ref. \cite{enrico} has also discussed issues related to this.
For completeness we include the corresponding matrices which
relate $\Delta_i$ to the chemical potential of the left-handed leptons, $\mu_{\nu_{i}}$, depending on which of the charged Yukawa couplings are in equilibrium as discussed in the main text of the paper. Note that the relationship between
$\mu_{\nu_{i}}$ and $L_i$ also varies depending on the corresponding charged Yukawa couplings that are in equilibrium.

When $\Gamma_{\tau} \gg H$, and sphaleron interactions are also in equilibrium, then

\bea
\left(\begin{array}{c}
\Delta_e \\
\Delta_{\mu} \\
\Delta_{\tau}
\end{array}\right) &= & \left(\begin{array}{ccc}
-\frac{22}{9} & -\frac{4}{9} & -\frac{4}{9}\\
-\frac{4}{9} & -\frac{22}{9} & -\frac{4}{9}\\
-\frac{4}{9} + \frac{5}{3}\frac{1}{3N+3} & -\frac{4}{9} +\frac{5}{3}\frac{1}{3N+3} & -\frac{4}{9}+ \frac{8}{3}\frac{1}{3N+2} -3
\end{array}\right)\left( \begin{array}{c}
\mu_{\nu_{e}} \\\mu_{\nu_{\mu}} \\ \mu_{\nu_{\tau}}
\end{array}\right),
\eea
where $N$ is the number of generations. For the range of temperatures we have focused on in the main part the relation between these quantities is modified to:

\bea
\left(\begin{array}{c}
\Delta_e \\
\Delta_{\mu} \\
\Delta_{\tau}
\end{array}\right) &= & \left(\begin{array}{ccc}
-\frac{22}{9} & -\frac{4}{9} & -\frac{4}{9}\\
-\frac{4}{9} + \frac{5}{3}\frac{1}{3N+4} & -\frac{4}{9}-3 + \frac{8}{3}\frac{1}{3N+4} & -\frac{4}{9} + \frac{8}{3}\frac{1}{3N+4}\\
-\frac{4}{9} + \frac{5}{3}\frac{1}{3N+4} & -\frac{4}{9} +\frac{8}{3}\frac{1}{3N+4} & -\frac{4}{9}-3+ \frac{8}{3}\frac{1}{3N+4} 
\end{array}\right)\left( \begin{array}{c}
\mu_{\nu_{e}} \\\mu_{\nu_{\mu}} \\ \mu_{\nu_{\tau}}
\end{array}\right).
\eea
Finally, when all charged Yukawa couplings are in equilibrium, the relation is simply

\beq
\Delta_i = \left(-\frac{4}{9} +\frac{8}{3}\frac{1}{4N+2}\right)\sum_j \mu_{\nu_{j}} - 3 \mu_{\nu_{i}}.
\eeq


\begin{thebibliography}{10}

\bibitem{FY} M.~Fukugita and T.~Yanagida. \newblock 
%Baryogenesis without grand unification. \newblock 
{\em Phys. Lett.}, B174:45, 1986.


\bibitem{sakharov} A.D. Sakharov. \newblock {\em JETP Lett.}, 5:24, 1967.
For a review, see A.~Riotto and M.~Trodden,
  %``Recent progress in baryogenesis,''
  Ann.\ Rev.\ Nucl.\ Part.\ Sci.\  {\bf 49}, 35 (1999)
  [arXiv:hep-ph/9901362].

\bibitem{kuzmin} V.A. Kuzmin, V.A. Rubakov, and M.E. Shaposhnikov. \newblock
{\em Phys. Lett.}, B155:36, 1985.

\bibitem{seesaw} 
P.~Minkowski,
%``Mu $\to$ E Gamma At A Rate Of One Out Of 1-Billion Muon Decays?,''
Phys.\ Lett.\ B {\bf 67} (1977) 421;
%%CITATION = PHLTA,B67,421;%%
M. Gell-Mann, P. Ramond and
R. Slansky,  {\em Proceedings of the Supergravity Stony Brook Workshop}, New
York 1979,  eds. P. Van Nieuwenhuizen and D. Freedman; T. Yanagida,  {\em
Proceedinds of the Workshop on Unified Theories and Baryon Number in the
Universe},  Tsukuba, Japan 1979, ed.s A. Sawada and A. Sugamoto;
R. N. Mohapatra, G. Senjanovic,
{\it Phys.Rev.Lett.} {\bf 44} (1980)912.
%

%\cite{Giudice:2003jh}
\bibitem{leptogen}
See, {\it e.g.}
G.~F.~Giudice, A.~Notari, M.~Raidal, A.~Riotto and A.~Strumia,
%``Towards a complete theory of thermal leptogenesis in the SM and MSSM,''
Nucl.\ Phys.\ B {\bf 685}, 89 (2004)
[arXiv:hep-ph/0310123].
%%CITATION = HEP-PH 0310123;%%
%\cite{Buchmuller:2004nz}
%\bibitem{Buchmuller:2004nz}
W.~Buchmuller, P.~Di Bari and M.~Plumacher,
%``Leptogenesis for pedestrians,''
Annals Phys.\  {\bf 315} (2005) 305
[arXiv:hep-ph/0401240].
%%CITATION = HEP-PH 0401240;%%

\bibitem{work}  A partial list:~W.~Buchmuller, P.~Di Bari and M.~Plumacher,
    Nucl.\ Phys.\ B {\bf 643} (2002) 367
[arXiv:hep-ph/0205349];
 W.~Buchmuller, P.~Di Bari and M.~Plumacher,
 %``A bound on neutrino masses from baryogenesis,''
 Phys.\ Lett.\ B {\bf 547} (2002) 128 [arXiv:hep-ph/0209301];
 G.~C.~Branco, R.~Gonzalez Felipe, F.~R.~Joaquim, I.~Masina,
   M.~N.~Rebelo and C.~A.~Savoy,
  %``Minimal scenarios for leptogenesis and CP violation,''
  Phys.\ Rev.\ D {\bf 67} (2003) 073025
  [arXiv:hep-ph/0211001];
J.~R.~Ellis, M.~Raidal and T.~Yanagida,
  %``Observable consequences of partially degenerate leptogenesis,''
  Phys.\ Lett.\ B {\bf 546} (2002) 228
  [arXiv:hep-ph/0206300];
 G.~C.~Branco, R.~Gonzalez Felipe, F.~R.~Joaquim and M.~N.~Rebelo,
  %``Leptogenesis, CP violation and neutrino data: What can we learn?,''
  Nucl.\ Phys.\ B {\bf 640} (2002) 202
  [arXiv:hep-ph/0202030];
 G.~C.~Branco, T.~Morozumi, B.~M.~Nobre and M.~N.~Rebelo,
   %``A bridge between CP violation at low energies and leptogenesis,''
   Nucl.\ Phys.\ B {\bf 617} (2001) 475
  [arXiv:hep-ph/0107164];
R.~N.~Mohapatra and S.~Nasri,
%``Leptogenesis and mu - tau symmetry,''
Phys.\ Rev.\ D {\bf 71} (2005) 033001
[arXiv:hep-ph/0410369];
R.~N.~Mohapatra, S.~Nasri and H.~B.~Yu,
%``Leptogenesis, mu - tau symmetry and theta(13),''
Phys.\ Lett.\ B {\bf 615} (2005) 231
[arXiv:hep-ph/0502026];
  A.~Broncano, M.~B.~Gavela and E.~Jenkins,
  %``Neutrino physics in the seesaw model,''
  Nucl.\ Phys.\ B {\bf 672} (2003) 163
  [arXiv:hep-ph/0307058]; Phys.\ Lett.\ B {\bf 552} (2003) 177
  [arXiv:hep-ph/0210271];A.~Pilaftsis,
%  %``CP violation and baryogenesis due to heavy Majorana neutrinos,''
  Phys.\ Rev.\ D {\bf 56} (1997) 5431
  [arXiv:hep-ph/9707235]; 
G.~C.~Branco, R.~Gonzalez Felipe, F.~R.~Joaquim and B.~M.~Nobre,
%%``Enlarging the window for radiative leptogenesis,''
arXiv:hep-ph/0507092. 


\bibitem{di}K.~Hamaguchi, H.~Murayama and T.~Yanagida,
%``Leptogenesis from sneutrino-dominated early universe,''
Phys.\ Rev.\ D {\bf 65} (2002) 043512
[arXiv:hep-ph/0109030].
%%CITATION = HEP-PH 0109030;%%


\bibitem{bound} S.~Davidson and A.~Ibarra,
  %``A lower bound on the right-handed neutrino mass from leptogenesis,''
  Phys.\ Lett.\ B {\bf 535} (2002) 25.


\bibitem{bdp2} W.~Buchmuller, P.~Di~Bari, and M.~Plumacher. %\newblock A bound
%on neutrino masses from baryogenesis. 
\newblock {\em Phys. Lett.},
B547:128--132, 2002.

\bibitem{HLNPS}
T.~Hambye, Y.~Lin, A.~Notari, M.~Papucci and A.~Strumia,
%``Constraints on neutrino masses from leptogenesis models,''
Nucl.\ Phys.\ B {\bf 695} (2004) 169
[arXiv:hep-ph/0312203].
%%CITATION = HEP-PH 0312203;%%

%\cite{Barbieri:1999ma}

\bibitem{barbieri}
R.~Barbieri, P.~Creminelli, A.~Strumia and N.~Tetradis,
%``Baryogenesis through leptogenesis,''
Nucl.\ Phys.\ B {\bf 575} (2000) 61
[arXiv:hep-ph/9911315].
%%CITATION = HEP-PH 9911315;%%



\bibitem{kor1}
T.~Endoh, T.~Morozumi and Z.~h.~Xiong,
  %``Primordial lepton family asymmetries in seesaw model,''
  Prog.\ Theor.\ Phys.\  {\bf 111}, 123 (2004)
[arXiv:hep-ph/0308276].

\bibitem{kor2}
T.~Fujihara, S.~Kaneko, S.~Kang, D.~Kimura, T.~Morozumi and M.~Tanimoto,
  %``Cosmological family asymmetry and CP violation,''
  Phys.\ Rev.\ D {\bf 72}, 016006 (2005)
[arXiv:hep-ph/0505076].


\bibitem{oscar} O. Vives, hep-ph/0512160.


%\cite{Pilaftsis:2005rv}
\bibitem{PU}
A.~Pilaftsis and T.~E.~J.~Underwood,
%``Electroweak-scale resonant leptogenesis,''
arXiv:hep-ph/0506107.
%%CITATION = HEP-PH 0506107;%%


\bibitem{todo} A. Abada, S. Davidson, F.X. Josse-Micheaux, M. Losada and
A. Riotto, to appear.


\bibitem{HT}
J.~A.~Harvey and M.~S.~Turner,
%``Cosmological Baryon And Lepton Number In The Presence Of Electroweak Fermion
%Number Violation,''A second important point which was pointed out in ref. \cite{barbieri} 
is the relationship between
the lepton asymmetry for each flavour $Y_{L_{i}}$ and the $B-L_i$ asymmetry.
Phys.\ Rev.\ D {\bf 42} (1990) 3344.
%%CITATION = PHRVA,D42,3344;%%


\bibitem{CDEO}
B.~A.~Campbell, S.~Davidson, J.~R.~Ellis and K.~A.~Olive,
%``On the baryon, lepton flavor and right-handed electron asymmetries of the
%universe,''
Phys.\ Lett.\ B {\bf 297} (1992) 118
[arXiv:hep-ph/9302221].
%%CITATION = HEP-PH 9302221;%%

\bibitem{CKO}
J.~M.~Cline, K.~Kainulainen and K.~A.~Olive,
%``Protecting the primordial baryon asymmetry from erasure by sphalerons,''
Phys.\ Rev.\ D {\bf 49} (1994) 6394
[arXiv:hep-ph/9401208].
%%CITATION = HEP-PH 9401208;%%


%\cite{Sigl:1992fn}
\bibitem{RS}
G.~Sigl and G.~Raffelt,
%``General kinetic description of relativistic mixed neutrinos,''
Nucl.\ Phys.\ B {\bf 406} (1993) 423.
%%CITATION = NUPHA,B406,423;%%

%\cite{Stodolsky:1986dx}
\bibitem{leo}
L.~Stodolsky,
%``On The Treatment Of Neutrino Oscillations In A Thermal Environment,''
Phys.\ Rev.\ D {\bf 36} (1987) 2273. \\
%%CITATION = PHRVA,D36,2273;%%
See also: G.~G.~Raffelt,
``Stars as laboratories for fundamental physics: The astrophysics of neutrinos,
axions, and other weakly interacting particles''.
%\href{http://www.slac.stanford.edu/spires/find/hep/www?irn=3668177}{SPIRES entry}

%\cite{McKellar:1992ja}
\bibitem{McKellar}
B.~H.~J.~McKellar and M.~J.~Thomson,
%``Oscillating doublet neutrinos in the early universe,''
Phys.\ Rev.\ D {\bf 49}, 2710 (1994).
%%CITATION = PHRVA,D49,2710;%%




%\cite{Raffelt:1992uj}
\bibitem{Raffelt:1992uj}
G.~Raffelt, G.~Sigl and L.~Stodolsky,
%``NonAbelian Boltzmann equation for mixing and decoherence,''
Phys.\ Rev.\ Lett.\  {\bf 70} (1993) 2363
[arXiv:hep-ph/9209276].
%%CITATION = HEP-PH 9209276;%%



%\cite{Enqvist:1990ad}
\bibitem{Enqvist:1990ad}
K.~Enqvist, K.~Kainulainen and J.~Maalampi,
%``Refraction And Oscillations Of Neutrinos In The Early Universe,''
Nucl.\ Phys.\ B {\bf 349} (1991) 754.
%%CITATION = NUPHA,B349,754;%%

%\cite{Foot:1996qc}
\bibitem{Foot:1996qc}
R.~Foot and R.~R.~Volkas,
% ``Studies of neutrino asymmetries generated by ordinary sterile neutrino
%oscillations in the early universe and implications for big bang
%nucleosynthesis bounds,''
Phys.\ Rev.\ D {\bf 55} (1997) 5147
[arXiv:hep-ph/9610229].
%%CITATION = HEP-PH 9610229;%%
\bibitem{ar} See, for instance, 
A.~Riotto,
  %``Towards a Nonequilibrium Quantum Field Theory Approach to Electroweak
  %Baryogenesis,''
  Phys.\ Rev.\ D {\bf 53}, 5834 (1996)
  [arXiv:hep-ph/9510271]; 
A.~Riotto,
  %``Supersymmetric electroweak baryogenesis, nonequilibrium field theory  and
  %quantum Boltzmann equations,''
  Nucl.\ Phys.\ B {\bf 518}, 339 (1998)
  [arXiv:hep-ph/9712221].

\bibitem{roulet} L.~Covi, E.~Roulet and F.~Vissani,
  %``CP violating decays in leptogenesis scenarios,''
  Phys.\ Lett.\ B {\bf 384} (1996) 169
  [arXiv:hep-ph/9605319].
\bibitem{Buchmuller:2003gz}
  W.~Buchmuller, P.~Di Bari and M.~Plumacher,
  %``The neutrino mass window for baryogenesis,''
  Nucl.\ Phys.\ B {\bf 665} (2003) 445
  [arXiv:hep-ph/0302092].
  %%CITATION = HEP-PH 0302092;%%


\bibitem{dip}
S.~Davidson,
%``From weak-scale observables to leptogenesis,''
JHEP {\bf 0303} (2003) 037
[arXiv:hep-ph/0302075].
%%CITATION = HEP-PH 0302075;%%



\bibitem{Fischler:1991gn}
M.~Fukugita and T.~Yanagida,
  %``Sphaleron Induced Baryon Number Nonconservation And A Constraint On
  %Majorana Neutrino Masses,''
  Phys.\ Rev.\ D {\bf 42}, 1285 (1990); 
 J.~A.~Harvey and M.~S.~Turner,
  %``Cosmological Baryon And Lepton Number In The Presence Of Electroweak
  %Fermion Number Violation,''
  Phys.\ Rev.\ D {\bf 42} (1990) 3344;  A.~E.~Nelson and S.~M.~Barr,
  %``Upper Bound On Baryogenesis Scale From Neutrino Masses,''
  Phys.\ Lett.\ B {\bf 246}, 141 (1990);
W.~Fischler, G.~F.~Giudice, R.~G.~Leigh and S.~Paban,
%``Constraints on the baryogenesis scale from neutrino masses,''
Phys.\ Lett.\ B {\bf 258} (1991) 45;
%%CITATION = PHLTA,B258,45;%%
W.~Buchmuller and T.~Yanagida,
%``Baryogenesis and the scale of B-L breaking,''
Phys.\ Lett.\ B {\bf 302} (1993) 240.
%%CITATION = PHLTA,B302,240;%%


\bibitem{DK}
S.~Davidson and R.~Kitano,
%``Leptogenesis and a Jarlskog invariant,''
JHEP {\bf 0403} (2004) 020
[arXiv:hep-ph/0312007].
%%CITATION = HEP-PH 0312007;%%


\bibitem{casasibarra} J.~A. Casas and A.~Ibarra. 
%\newblock Oscillating
%neutrinos and mu $\to$ e, gamma. 
\newblock {\em Nucl. Phys.}, B618:171--204,
2001.


%\cite{cosmo}
\bibitem{cosmo}
D.~N.~Spergel {\it et al.}  [WMAP Collaboration],
%``First Year Wilkinson Microwave Anisotropy Probe (WMAP) Observations:
%Determination of Cosmological Parameters,''
Astrophys.\ J.\ Suppl.\  {\bf 148} (2003) 175
[arXiv:astro-ph/0302209].
%%%CITATION = ASTRO-PH 0302209;%%
P.~Crotty, J.~Lesgourgues and S.~Pastor,
%``Current cosmological bounds on neutrino masses and relativistic relics,''
Phys.\ Rev.\ D {\bf 69} (2004) 123007
[arXiv:hep-ph/0402049];
%%CITATION = HEP-PH 0402049;%%
U.~Seljak {\it et al.},
% ``Cosmological parameter analysis including SDSS Ly-alpha forest and  galaxy
%%bias: Constraints on the primordial spectrum of fluctuations,  neutrino mass,
%%and dark energy,''
Phys.\ Rev.\ D {\bf 71} (2005) 103515
[arXiv:astro-ph/0407372].
%%CITATION = ASTRO-PH 0407372;%%


\bibitem{marta}
A.~Abada and M.~Losada,
%``Leptogenesis with four gauge singlets,''
Nucl.\ Phys.\ B {\bf 673} (2003) 319
[arXiv:hep-ph/0306180].
%%CITATION = HEP-PH 0306180;%%
L.~Boubekeur, T.~Hambye and G.~Senjanovic,
%``Low-scale leptogenesis and soft supersymmetry breaking,''
Phys.\ Rev.\ Lett.\  {\bf 93} (2004) 111601
[arXiv:hep-ph/0404038].
%%CITATION = HEP-PH 0404038;%%
A.~Abada, H.~Aissaoui and M.~Losada,
%``A model for leptogenesis at the TeV scale,''
Nucl.\ Phys.\ B {\bf 728} (2005) 55
[arXiv:hep-ph/0409343].
%%CITATION = HEP-PH 0409343;%%
 M.~Senami and K.~Yamamoto,
%  %``Leptogenesis with supersymmetric Higgs triplets in TeV region,''
  arXiv:hep-ph/0305202.
  M.~Senami and K.~Yamamoto,
%  %``Lepton flavor violation with supersymmetric Higgs triplets in TeV  region
%  %for neutrino masses and leptogenesis,''
 Phys.\ Rev.\ D {\bf 69} (2004) 035004
  [arXiv:hep-ph/0305203].
 L.~Boubekeur,
%  %``Leptogenesis at low scale,''
  arXiv:hep-ph/0208003.



\bibitem{gravitino}
J.~R.~Ellis, J.~E.~Kim and D.~V.~Nanopoulos,
  %``Cosmological Gravitino Regeneration And Decay,''
  Phys.\ Lett.\ B {\bf 145}, 181 (1984).

\bibitem{KW}
E.~W.~Kolb and S.~Wolfram,
%``Baryon Number Generation In The Early Universe,''
Nucl.\ Phys.\ B {\bf 172} (1980) 224
[Erratum-ibid.\ B {\bf 195} (1982) 542].
%%CITATION = NUPHA,B172,224;%%

\bibitem{enrico}
E. Nardi, Y. Nir, J. Racker and E. Roulet, hep-ph/0512052.

\bibitem{nnrr} E. Nardi, Y. Nir, J. Racker and E. Roulet, hep-ph/0601084.
\end{thebibliography}
\end{document}